\documentclass{aa}

\usepackage{graphicx}
\usepackage{natbib}
\usepackage{lscape}
\usepackage{longtable}
\usepackage{color}
\usepackage{txfonts}
\begin{document}

   \title{Stellar parameters of early M dwarfs from ratios of spectral features
          at optical wavelengths
   \thanks{
     Based on data products from observations made with ESO Telescopes at the La Silla Paranal Observatory
     under programmes ID 072.C-0488(E), 082.C-0718(B), 085.C-0019(A), 180.C-0886(A), 183.C-0437(A),
     and 191.C-0505(A), as well as data from the Italian Telescopio Nazionale Galileo (TNG) Archive
     (programmes ID CAT-147, and A27CAT\_83). 
     }\fnmsep
   \thanks{
     Our computational codes including the full and more detailed version of
     Tables~\ref{examples_of_teff_ratios},~\ref{calibrations_sptype}
     and ~\ref{calibrations_metallicity} are available at
     http://www.astropa.inaf.it/\textasciitilde{}jmaldonado/Msdlines.html}   
     }


   \author{J. Maldonado 
          \inst{1}
          \and  L. Affer
          \inst{1}
          \and  G. Micela
          \inst{1}
          \and G. Scandariato
          \inst{2}   
          \and M. Damasso
          \inst{3} 
          \and B. Stelzer
          \inst{1}  
          \and M. Barbieri
          \inst{4}  
          \and L. R. Bedin
          \inst{4} 
          \and K. Biazzo
          \inst{2}
          \and A. Bignamini 
          \inst{5}
          \and F. Borsa
          \inst{6} 
          \and R.U. Claudi
          \inst{4}  
          \and E. Covino
          \inst{7}
          \and S. Desidera
          \inst{4}
          \and M. Esposito 
          \inst{8,9} 
          \and R. Gratton
          \inst{4}
          \and J. I. Gonz\'alez Hern\'andez   
          \inst{8,9}
          \and A.F. Lanza
          \inst{2} 
          \and A. Maggio
          \inst{1}   
          \and E. Molinari
          \inst{10,11} 
          \and I. Pagano
          \inst{2}  
          \and M. Perger
          \inst{12}  
          \and I. Pillitteri
          \inst{1}  
          \and {G. Piotto}
          \inst{13,4}
          \and E. Poretti
          \inst{6}
          \and L. Prisinzano
          \inst{1}
          \and R. Rebolo
          \inst{8,9}
          \and I. Ribas  
          \inst{12} 
          \and E. Shkolnik
          \inst{14} 
          \and J. Southworth
          \inst{15} 
          \and A. Sozzetti
          \inst{3}
          \and A. Su\'arez Mascare\~no
          \inst{8,9} 
          }  

             \institute{INAF - Osservatorio Astronomico di Palermo,
              Piazza Parlamento 1, I-90134 Palermo, Italy
             \and  
              INAF - Osservatorio Astrofisico di Catania, Via S. Sofia 78, I-95123, Catania, Italy
             \and  
              INAF - Osservatorio Astrofisico di Torino, Via Osservatorio 20, I-10025, Pino Torinese, Italy
             \and  
              INAF - Osservatorio Astronomico di Padova, Vicolo Osservatorio 5, I-35122, Padova, Italy
             \and  
              INAF - Osservatorio Astronomico di Trieste, via Tiepolo 11, 34143 Trieste, Italy  
             \and  
              INAF - Osservatorio Astronomico di Brera, Via E. Bianchi 46, I-23807 Merate (LC), Italy
             \and  
              INAF – Osservatorio Astronomico di Capodimonte, via Moiariello, 16, 80131 Naples, Italy       
             \and  
              Instituto de Astrof\'isica de Canarias, E-38205 La Laguna, Tenerife, Spain   
             \and 
              Universidad de La Laguna, Dpto. Astrof\'isica, E-38206 La Laguna, Tenerife, Spain
             \and  
              Fundaci\'on Galileo Galilei - INAF, Rambla Jos\'e Ana Fernandez P\'erez 7, E-38712 Bre\~{n}a Baja, TF - Spain
             \and  
              INAF - IASF Milano, via Bassini 15, I-20133 Milano, Italy
             \and  
              Institut de Ci\'encies de l'Espai (CSIC-IEEC), Campus UAB, Facultat de Ci\'encies, Torre C5 parell, 2a
              planta, E-08193 Bellaterra, Spain
             \and  
              Dip. di Fisica e Astronomia Galileo Galilei – Universit\`a di Padova,
              Vicolo dell'Osservatorio 2, I-35122, Padova, Italy
             \and  
              Lowell Observatory, 1400 W. Mars Hill Road, Flagstaff, AZ, 86001, USA
             \and  
              Astrophysics Group, Keele University, Staffordshire, ST5 5BG, UK
             }

   \offprints{J. Maldonado \\ \email{jmaldonado@astropa.inaf.it}}
   \date{Received ...; Accepted ....}

 
  \abstract
   {
    Low-mass stars have been recognised as promising targets in the search for rocky,
    small planets with the potential of supporting life. As a consequence, Doppler
    search programmes using high-resolution spectrographs like HARPS or HARPS-N
    are providing huge quantities of optical spectra of M dwarfs. However, determining the
    stellar parameters of M dwarfs using optical spectra has proven
    to be challenging.    
   }
   {
    We aim to calibrate empirical relationships  
    to determine
    accurate stellar parameters for early M dwarfs (spectral types M0-M4.5) using the same spectra
    that are used
    for the radial velocity determinations, without the necessity of
    acquiring IR spectra or relying on atmospheric models and/or photometric calibrations.    
   }
   {
    Our methodology consists in the use of ratios of pseudo equivalent widths of spectral
    features as a temperature diagnostic, a technique largely used in solar-type stars. 
    Stars with effective temperatures obtained from interferometric estimates of their radii
    are used as calibrators. Empirical calibrations for the spectral type are also provided.
    Combinations of features and ratios of features are used to derive calibrations for
    the stellar metallicity. 
    Our methods are then applied to a large sample of M dwarfs that are currently being
    observed in the framework of the HARPS GTO  search for extrasolar planets.
    The derived temperatures
    and metallicities are used  together with photometric estimates of mass, radius, and
    surface gravity to calibrate empirical relationships for these parameters. 
    }  
   {
    A large list of spectral features in the optical spectra of early M dwarfs was identified.
    From this list the pseudo equivalent width of roughly  43\% of the features
    shows a strong anticorrelation with the effective temperature. The correlation with the stellar
    metallicity is weaker. A total of  112 temperature sensitive ratios have been identified and calibrated
    over the range  3100-3950 K, providing effective temperatures with typical uncertainties of the order 
    of 70 K. Eighty-two ratios of pseudo equivalent widths of features were calibrated to derive spectral types
    within 0.5 subtypes for stars with spectral types between K7V and M4.5V.  
    Regarding stellar metallicity,  696 combinations of pseudo equivalent widths of individual features and
    temperature-sensitive ratios have been calibrated, over the metallicity range from -0.54 to +0.24 dex,
    with estimated uncertainties in the
    range of 0.07-0.10 dex.
    We provide our own empirical calibrations for  stellar mass, radius, and surface gravity.
    These parameters are found  to show a dependence on the stellar metallicity.
    For a given effective temperature, lower metallicities
    predict lower masses and radii, as well as larger gravities.    
   }
    {
     }

  \keywords{techniques: spectroscopic -stars: late-type -stars: low-mass -stars: fundamental parameters}

  \maketitle

\section{Introduction}
\label{introduccion}

 Ratios of equivalent widths or central depths of absorption lines with
 different excitation potentials have been widely
 used as temperature indicators in different kind of stars including solar-type
 \citep[e.g.][]{1991PASP..103..439G,1994PASP..106.1248G,2002A&A...386..286C,2003A&A...411..559K,
 2007AN....328..938B,
 2007HiA....14..598M,2010A&A...512A..13S,2012MNRAS.426..484D,2014MNRAS.439.1028D,2014arXiv1412.8168D}, 
 giant stars \citep[e.g][]{1989ApJ...347.1021G,2000AN....321..277S}, 
 and supergiants \citep[e.g.][]{2000A&A...358..587K}. 
 To the best of our knowledge, this technique has however not been extended
 to the low-mass stars regime most likely due to the difficulties in analysing
 their optical spectrum, mainly covered by molecular
 bands (in particular TiO and water) which blend or hide
 most of the atomic lines commonly used in the spectral
 analysis of solar-type stars. 
 Furthermore, M dwarfs are intrinsically faint in 
  the optical.

 The accurate determination of stellar parameters for M dwarfs has
 proven to be a difficult task.
 Regarding stellar metallicity, some photometric calibrations based on optical and near-infrared photometry exist,
 a technique pioneered by \cite{2005A&A...442..635B} and updated
 by \cite{2009ApJ...699..933J,2010A&A...519A.105S} and more recently by
 \citet[][hereafter NE12]{2012A&A...538A..25N},
 although they require accurate parallaxes
 and magnitudes which are
 usually available only for nearby and bright stars. 
 Another common technique to characterise M dwarf metallicities 
 is based on the use of spectroscopic indices which measure the 
 relative strength of the TiO molecular band 
 with respect to the CaH molecular bands near 7000\AA \space
 \citep{2007ApJ...669.1235L,2012AJ....143...67D,2013AJ....145..102L}. 
 Since the M dwarfs spectral energy
 distribution peaks at infrared wavelengths,
 some previous works have performed a search for spectral features and indices in
 this spectral region.
 In particular, \cite{2012ApJ...748...93R}
 use the equivalent width of the Na~{\sc i} and Ca~{\sc i} triplet and the
 H$_{\rm 2}$O-K2 index in the K band of the spectra. 
 This methodology has been also applied by
 \cite{2012ApJ...747L..38T}  and \cite{2013AJ....145...52M}  providing calibrations for 
 H and J/optical spectral bands, respectively.
 Large samples of M dwarfs have been characterised by means
 of near-infrared indices in the recent works by
 \cite{2014AJ....147...20N} and \cite{2014MNRAS.443.2561G}.
 On the other hand, spectral synthesis has been tested in several works,
 usually on small number of stars focusing mainly on strong atomic lines
 or on spectral windows known to be less affected by molecular lines
 \citep{2005MNRAS.356..963W,2006ApJ...653L..65B,2007PASP..119...90M,
  2012A&A...542A..33O,2014A&A...564A..90R}.  

 Concerning the effective temperature, very few M dwarfs are bright enough
 for a direct measurement of their radii 
 \citep[e.g.][hereafter BO12]{2012ApJ...757..112B}, a technique
 pioneered by \cite{2003A&A...397L...5S}.
 The most common technique for determining the effective temperature 
 of an M dwarf is the comparison
 of observed spectra
 with models atmosphere \citep[e.g.][]{2013AJ....145..102L,2014MNRAS.443.2561G}.
 \citet[][hereafter CA08]{2008MNRAS.389..585C} 
 provides optical/near-infrared photometric calibrations 
 based on an extension of the infrared flux method (IRFM)
 for  FGK dwarfs from \cite{2006MNRAS.373...13C}  to M dwarfs.
 However, significant systematic differences between temperatures
 based on CA08 calibrations and temperatures based on interferometric
 radii measurements have been recently noted by \cite{2015arXiv150101635M}.

 Despite the intrinsic difficulties in their characterisation, low-mass
 stars are nowadays at the centre 
 in the search for small, rocky planets with the potential capability
 of hosting life
 \citep[e.g.][]{2013ApJ...767...95D,2013EPJWC..4703006S} 
 In particular, the radial velocity searches currently ongoing with HARPS
 at La Silla,
 and in the framework of the {\it Global Architecture of Planetary Systems}
 project\footnote{http://www.oact.inaf.it/exoit/EXO-IT/Projects/Entries/2011/12/\\27\_GAPS.html}
 \citep[GAPS;][]{2013A&A...554A..28C} 
 at the Telescopio Nazionale Galileo (TNG) with HARPS-N,
 are producing a large quantity
 of high resolution and high signal-to-noise ratio spectra.
 Exoplanet searches would certainly
 benefit from a methodology to determine accurate stellar parameters using
 the same spectra that are being used for the radial velocity determinations,
 i.e., without the necessity of observing at infrared facilities
 (usually from space)
 or relying on
 atmospheric models. 
 Following this line of reasoning,   
 a methodology to characterise M dwarfs from high resolution optical spectra by using
 pseudo equivalent widths of features has been presented in a recent work 
 by \citet[][hereafter NE14]{2014A&A...568A.121N}. 

 The idea of pseudo equivalent width can be further exploited
 in order to calibrate empirical relationships for M dwarfs. 
 This is the goal of this paper, in which we present a large database of
 empirical calibrations of spectral features, ratios of features and combinations
 of both with the  aim of deriving T$_{\rm eff}$, spectral type, and metallicity,
 for early M dwarf stars by using optical
 HARPS and HARPS-N spectra (wavelength range 383-693 nm).
 Unlike NE14 we take as reference temperature scale the one provided
 by BO12 and not CA08. We also apply our method to derive spectral types. 
 Furthermore, we provide relations for stellar masses, radii, and surface gravities
 so these quantities can be obtained without using parallaxes or photometry.
 We use our methods to characterise
 in an homogeneous and coherent way a sample of the M dwarfs which
 are currently being monitored in the HARPS GTO radial velocity programme
 \citep{2013A&A...549A.109B}.
 Late M stars are excluded from this study since exoplanet search around
 these stars is difficult at optical wavelengths.

 This paper is organised as follows.
 Section~\ref{spectroscopic_data} describes the spectroscopic
 data used in this work.
 Section~\ref{methodology} 
 describes our methodology and how empirical calibrations for
 the main stellar parameters are built. These calibrations are then
 applied to a large sample of stars and results are compared
 with other techniques in Section~\ref{comparison}.
 The derived temperatures and metallicities are used to build
 empirical calibrations for stellar masses, radii, and gravities
 in Section~\ref{mass_radii_gravity}.
 Our conclusions follow in Section~\ref{summary}.

\section{Spectroscopic data}
\label{spectroscopic_data}

  This work makes use of HARPS \citep{2003Msngr.114...20M} and
  HARPS-N \citep{2012SPIE.8446E..1VC}
  spectra mostly taken
  from archive. Specifically, the data is taken from:
  {\it i)} The ESO pipeline processed FEROS and HARPS archive
  \footnote{http://archive.eso.org/wdb/wdb/eso/repro/form};
  {\it ii)} The ESO Science Data products Archive
  \footnote{http://archive.eso.org/wdb/wdb/adp/phase3\_spectral/form?phase3\_collection
  =HARPS}; and
  {\it iii)} The TNG Archive
  \footnote{http://ia2.oats.inaf.it/archives/tng}.
  The corresponding ESO and TNG programme IDs are 
  listed in the footnote to the paper title.  
  In addition to the data from archive,
  some HARPS-N spectra have been provided by the GAPS team.

  The instrumental setup of HARPS and HARPS-N is almost identical so
  data from both spectrographs can be used together. 
  The spectra cover the range 383-693 nm (HARPS-N), and
  378-691 nm (HARPS). Both instruments provide a resolving power
  of R$\sim$ 115000. The spectra are provided already reduced
  using ESO/HARPS-N standard calibration pipelines.
  Typical values of the signal-to-noise ratio are between
  50 and 90 (measured at $\sim$ 5500 \space \AA). Wavelengths are on air. 
  The spectra were corrected from radial
  velocity shifts by using the IRAF
  \footnote{IRAF is distributed  by the National Optical Astronomy Observatory, which
  is operated by the Association of Universities for Research in Astronomy, Inc., under
  contract with the National Science Foundation.}
  task {\it dopcor}. For this purpose, we used the accurate radial velocities (measured by applying
  the cross-correlation technique)
  which are provided with
  the reduced spectra.

\section{Methodology}
\label{methodology}

\begin{figure}[!htb]
\centering
\includegraphics[angle=270,scale=0.45]{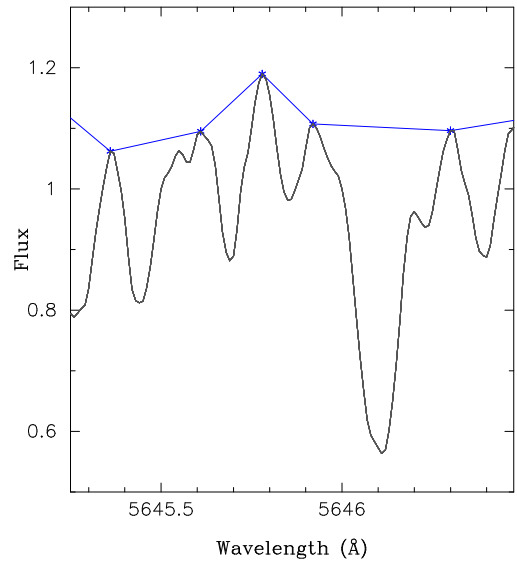}
\caption{Example of the measurement of EWs. 
The figure shows a portion of the spectrum of the star Gl 49.
The flux between the peaks of the features is measured over the blue lines.
}
\label{examples_pseudo_ews_measurements}
\end{figure}

  The optical spectrum of an M dwarf is a forest of lines 
  and molecular bands usually heavily blended in which identifying
  individual lines (of moderate strength) or measure equivalent widths
  is a difficult task. 
  In order to overcome this limitation we follow the idea depicted
  by NE14 and instead of considering equivalent
  widths or depths of lines, we consider pseudo equivalent
  widths (hereafter EWs) of features. A feature can be a line or a blend
  of lines. The pseudo equivalent width is defined as the traditional 
  equivalent width, with the 
  difference that it is not measured with respect to a continuum
  normalised to unit,
  but to the value of the flux between the peaks of the feature
  at each wavelength:  

  \begin{equation}
  EW = \sum\frac{F_{pp}-F_{\lambda}}{F_{pp}}\Delta \lambda
  \end{equation}

  \noindent  where F$_{\lambda}$ is the stellar flux, F$_{\rm pp}$ denotes the
  value of flux between the peaks of
  the feature at each integration step, and $\Delta\lambda$
  is the spectra wavelength's step.
  An estimate of the uncertainty
  in the measured EWs is given by:

  \begin{equation}
  \sigma_{EW} = \frac{\Delta \lambda}{<F_{pp}>}\sigma_{F_{pp}} 
  \end{equation}

  \noindent where $<F_{\rm pp}>$ is the mean value of the flux between the peaks of
  the feature, and $\sigma_{F_{pp}}$ its corresponding standard deviation.
  Figure~\ref{examples_pseudo_ews_measurements} illustrates how the EWs are measured. 

  An initial list of 4224 features was built taking as reference the 
  spectra of the star Gl 49, a M1.5V low-mass dwarf. 
  Spectroscopic observations of this star
  were carried out within the framework of the GAPS project with HARPS-N. 
  Only the red region of the spectra (5300 - 6900\AA) was 
  considered since the blue part of the optical spectrum of an M dwarf usually
  suffers from
  lower signal-to-noise ratio. Regions of the spectra affected by chromospheric
  activity  and atmospheric absorption were avoided.

\begin{figure*}[!htb]
\centering
\begin{minipage}{0.48\linewidth}
\includegraphics[angle=270,scale=0.55]{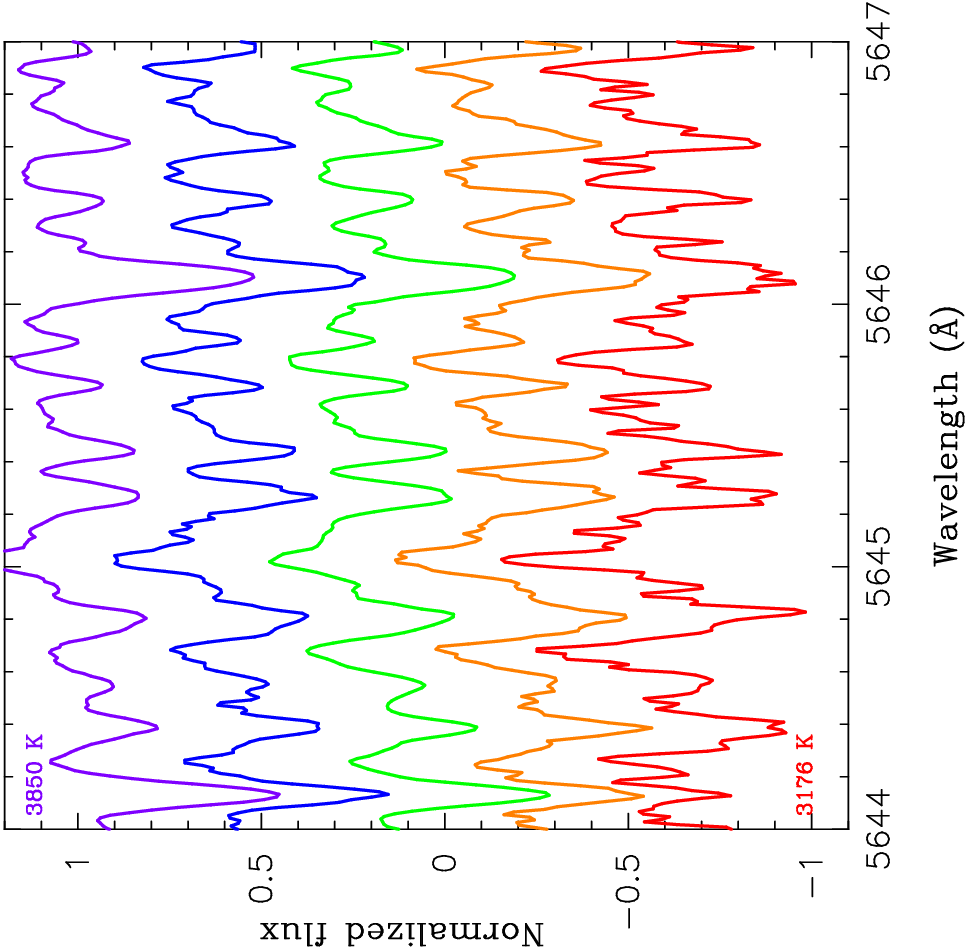}
\end{minipage}
\begin{minipage}{0.48\linewidth}
\includegraphics[angle=270,scale=0.55]{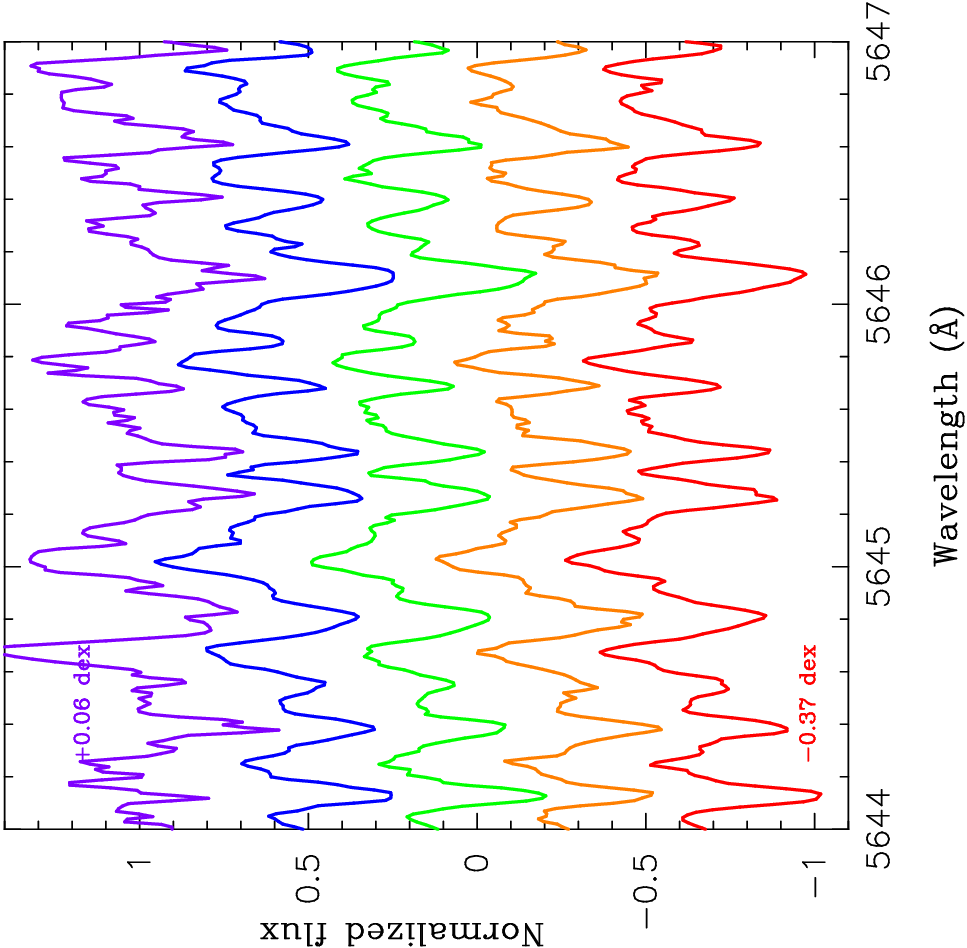}
\end{minipage}
\caption{
 Left: Portion of the stellar spectrum of several T$_{\rm eff}$ calibrators 
 (see text for details)
 with similar metallicity
 (from -0.07 to + 0.15 dex) but different effective temperature:
 3850 K (purple), 3701 K (blue), 3646 K (green), 3520 K (orange), and 3176 K (red).
 Right:  Portion of the stellar spectrum of several metallicity calibrators 
 (see text for details)
 with similar effective
 temperature
 (3450 - 3550 K) but different metallicity:
 +0.06 dex (purple), +0.01 dex (blue), -0.11 dex (green), -0.20 dex (orange), and -0.37 dex (red).
 For the sake of clarity, an offset of -0.40 in flux was applied between the spectra.}
\label{effect_of_teff_and_metal}
\end{figure*}

  Figure~\ref{effect_of_teff_and_metal} shows the effects of effective temperature
  and metallicity on the EWs measurement.
  In the left panel a portion of the stellar spectra is shown
  for stars with similar metallicities but different
  T$_{\rm eff}$. The same portion of the spectra is shown
  in the right panel, this time for stars of similar
  T$_{\rm eff}$ but different metallicities.

  A list of calibrators was built for each
  of the basic stellar parameters considered in the
  present work 
  (T$_{\rm eff}$, spectral type, and metallicity), following
  different criteria
  as explained in the next subsections.

\subsection{Effective temperature}

\begin{figure*}[!htb]
\centering
\begin{minipage}{0.48\linewidth}
\includegraphics[angle=270,scale=0.55]{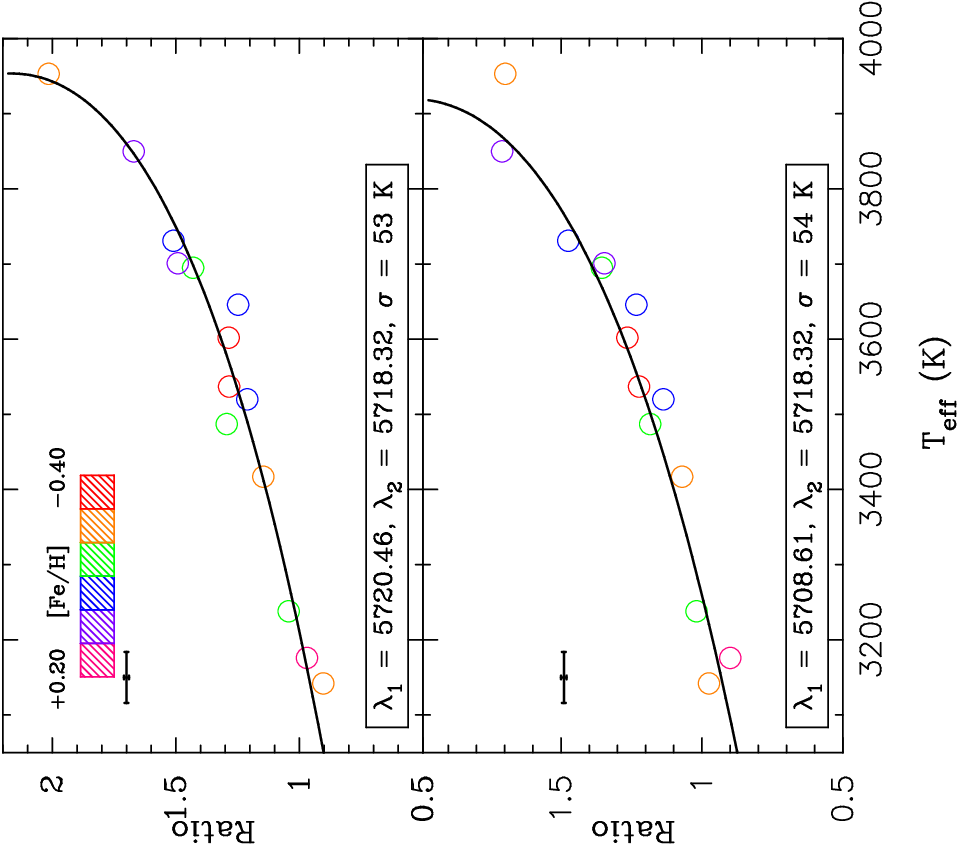}
\end{minipage}
\begin{minipage}{0.48\linewidth}
\includegraphics[angle=270,scale=0.55]{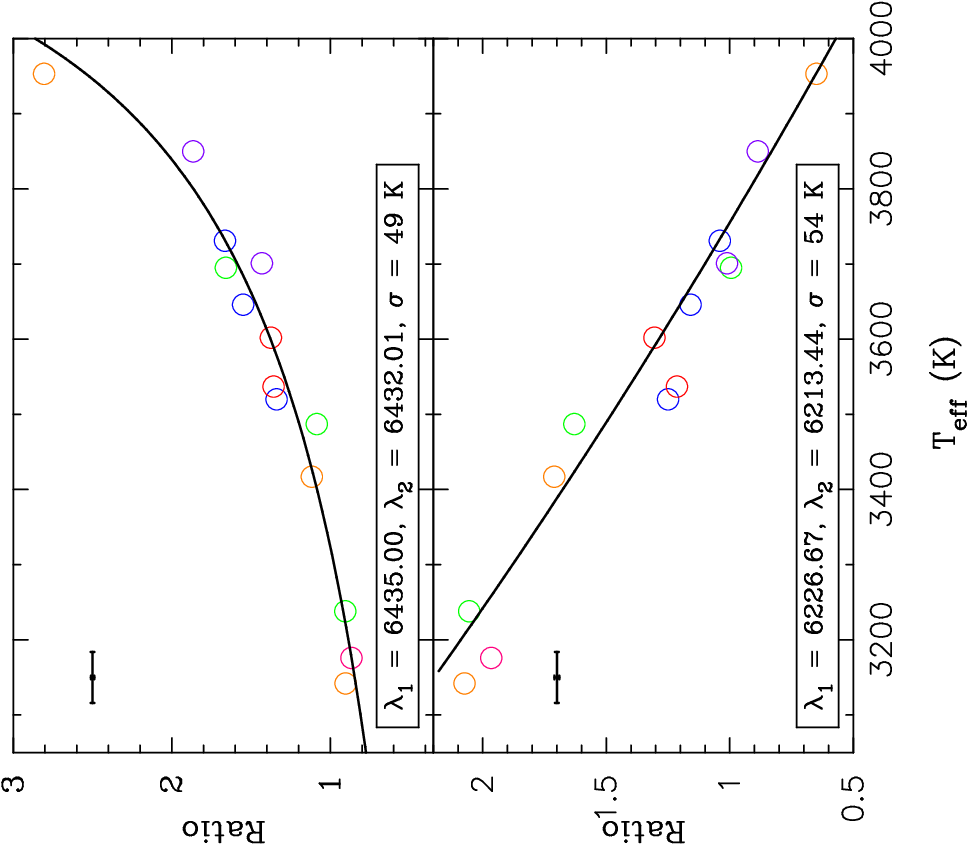}
\end{minipage}
\caption{
Examples of ratios of some features identified to be
sensitive to T$_{\rm eff}$ in early-M dwarfs.
Stars are plotted using different colours according to their metallicity
(using 0.10 dex length bins).
Median uncertainties are shown in the left upper corner of each plot. 
The corresponding fits are also plotted.
The features' central wavelengths as well as the rms standard
deviation of the residuals are given in each plot.
}
\label{examples_of_teff_ratios}
\end{figure*}

  The accuracy of the T$_{\rm eff}$ derived from temperature sensitive ratios is intimately 
  tied to the accuracy of the temperature of the stars used as calibrators.
  We used as calibrators the sample of early M dwarfs with angular sizes 
  obtained with long-baseline interferometry to better than
  5\% given in BO12.

  BO12
  list 22 low-mass stars with spectral-types
  equal to or cooler than K5V, spanning a range of T$_{\rm eff}$ between
  3000 and 4500 K. 
  HARPS spectra were obtained for  seven stars from the ESO archive,
  whilst HARPS-N spectra were obtained for 
  three stars from the TNG archive.
  HARPS-N spectra for two further stars have been provided
  by the GAPS team. 
   To these stars, we added two more from the recent work
   by \cite{2014MNRAS.438.2413V} who analyze their stars in the same way as BO12
  \footnote{The authors list three M dwarfs but we were unable
   to obtain a HARPS or HARPS-N spectra of Gl 649.}.
   
   \cite{2013ApJ...779..188M} revised the temperature scale of BO12 by noticing
   an issue regarding the determination of the bolometric flux of the stars. 
   Although the temperature differences are relatively small, we use the
   set of updated temperatures. For the two stars taken from 
   \cite{2014MNRAS.438.2413V}, updated T$_{\rm eff}$ values computed in the
   same way as in \cite{2013ApJ...779..188M} are provided in
   \cite{2014arXiv1412.2758N}.
    The final list of T$_{\rm eff}$ calibrators amounts to  14 stars whose
    parameters are listed in Table~\ref{calibration_sample}.

\begin{table}
\centering
\caption{The effective temperature calibration sample.}
\label{calibration_sample}
\begin{tabular}{lcc}
\hline\noalign{\smallskip}
 Star   &        SpT       & T$_{\rm eff}$   \\ 
        &                  &  (K)          \\
\hline
GJ338A   &  M0V      &  3953 $\pm$   41  \\
GJ205    &  M1.5V    &  3850 $\pm$   22  \\
GJ880    &  M1.5V    &  3731 $\pm$   16  \\
GJ176    &  M2.5V    &  3701 $\pm$   90  \\ 
GJ887    &  M0.5V    &  3695 $\pm$   35  \\
GJ526    &  M1.5V    &  3646 $\pm$   34  \\
GJ15A    &  M1.5V    &  3602 $\pm$   13  \\
GJ412A   &  M1V      &  3537 $\pm$   41  \\
GJ436    &  M3V      &  3520 $\pm$   66  \\
GJ581    &  M2.5V    &  3487 $\pm$   62  \\ 
GJ725A   &  M3V      &  3417 $\pm$   17  \\
GJ699    &  M4V      &  3238 $\pm$   11  \\
GJ876    &  M5V      &  3176 $\pm$   20  \\
GJ725B   &  M3.5V    &  3142 $\pm$   29  \\
\hline
\end{tabular}
\tablefoot{ 
Effective temperatures are from \cite{2013ApJ...779..188M}, and spectral types
for BO12. For GJ176 and GJ876, T$_{\rm eff}$ values are from \cite{2014arXiv1412.2758N},
and spectral types from the GJ catalogue \citep{1991adc..rept.....G}.
}
\end{table}

  Starting from our initial list of 4224 identified features, the 
  EW of all features were measured in all calibration stars.
  In order to avoid possible dependencies on microturbulence, rotation,
  or stellar metallicity, we rejected features with  
  EW $<$ 20 m\AA \space or EW $>$ 120 m\AA \space in any of the
  calibration stars, thus excluding too weak and too strong features.
  We also rejected lines with relative errors
  ($\frac{\sigma_{EW}}{EW}$) larger than 2\%. 
  For every possible ratio of features a Spearman's
  correlation test was computed to check whether the ratio is temperature
  sensitive or not.
  The ratios were selected with the only condition that the features
  are separated by no more than  15\AA.
  This limit was set in order to 
  avoid problems with scattered light correction or continuum normalisation.
  All ratios with a probability of correlation by chance
  lower than 2\% were considered for further study
  \footnote{
  Ideally, for a ratio of lines to be temperature-sensitive the excitation
  potential of the lines, $\chi$, must differ as much
  as possible. This is because the EWs of lines with higher $\chi$
  change with T$_{\rm eff}$ faster than those of lines 
  with lower  $\chi$ values \citep{1994PASP..106.1248G}.}. 

  Following \cite{2003A&A...411..559K} for each considered EW ratio we 
  fitted the T$_{\rm eff}$-ratio relationship to several functions:
  a Hoerl function (T$_{\rm eff}$ = ab$^{\rm r}$$\times$r$^{\rm c}$), a modified
  Hoerl function (T$_{\rm eff}$ = ab$^{\rm 1/r}$$\times$r$^{\rm c}$),
  a power-law function (T$_{\rm eff}$ = a$\times$r$^{\rm b}$),
  an exponential law function T$_{\rm eff}$ = a$\times$b$^{\rm r}$, and a
  logarithmic function T$_{\rm eff}$ = a + b$\times$$\ln$(r), where
  r = EW$_{\rm 1}$/EW$_{\rm 2}$ is the ratio between the EW of two features.  
  All fits were performed using a nonlinear least-squares fitting routine in IDL
  \citep[MPFIT;][]{2009ASPC..411..251M} taking into account the uncertainties
  in T$_{\rm eff}$.
  For each calibration we selected the function that produced the
  smallest standard deviation, retaining only those calibrations with 
  a standard deviation smaller than 75 K. 
  The number of selected
  temperature-sensitive ratios amounts to  112.

  Given our relatively low number of calibrators we performed an additional
  check to ensure that the selected ratios are not correlated with T$_{\rm eff}$ 
  simply by coincidence. We created 1000 series of simulated random effective
  temperatures and errors, keeping the media and the standard deviation of the
  original data. For each series of simulated data we repeted our analysis
  and computed the number of ``suitable'' calibrations. 
  The results show that in 98\% of the simulations our methodology does not recover any suitable
  T$_{\rm eff}$-calibration, whilst only in 0.8\% of the simulations the number
  of obtained calibrations is larger than 10. 
  We therefore conclude that it is very unlikely
  that our obtained T$_{\rm eff}$-ratios are correlated with T$_{\rm eff}$
  just by chance. 
  
  Some examples  of the selected
  temperature-sensitive ratios are shown in Figure~\ref{examples_of_teff_ratios},
  whilst full details regarding the calibrations for the same examples
  can be found in Table~\ref{calibrations_data}. 

\begin{table}
\centering
\caption{
Coefficients of our feature ratio-temperature relations. Columns (1) to (4) provide information
about the features involved in the ratio (central wavelength and width in \AA), 
while columns (5) to (9) show the coefficients of the best-fitting relationships, their functional
form, and the corresponding standard deviation of the T$_{\rm eff}$ calibration.
Only four examples are shown here. The same examples are shown in 
Figure~\ref{examples_of_teff_ratios}.
}
\label{calibrations_data}
\begin{scriptsize}
\begin{tabular}{ccccccccc}
\hline\noalign{\smallskip}
$\lambda_{\rm 1}$ & $\Delta\lambda_{\rm 1}$ & $\lambda_{\rm 2}$ & $\Delta\lambda_{\rm 2}$ &
  a  &  b  & c &  func.$^{\dag}$ & $\sigma$(K) \\
(1) & (2) & (3) & (4) & (5) & (6) & (7) & (8) & (9) \\
\hline
6435.00 &  0.38  & 6432.01 & 0.34  & 4529.63 &  0.73   & -0.015 & MH & 49 \\
5720.46 &  0.33  & 5718.32 & 0.28  & 4842.67 &  0.66   &  0.89  & H  & 53 \\
6226.67 &  0.35  & 6213.44 & 0.29  & 4350.09 &  0.86   &    -   & E  & 54 \\
5708.61 &  0.30  & 5718.32 & 0.28  & 5122.92 &  0.634  &  0.92  & H  & 54 \\
\hline
\multicolumn{9}{l}{$^{\dag}$: H: Hoerl; MH: modified Hoerl; PL: Power-law; E: Exponential; L: Logarithmic}\\
\end{tabular}
\end{scriptsize}
\end{table}

\subsection{Spectral Types}
\label{spectraltypes}

 A similar approach was followed to derive spectral types.
 HARPS or HARPS-N spectra were obtained for
 a sample of 33 stars with homogeneously derived spectral
 types in \cite{1991ApJS...77..417K} and \cite{1994AJ....108.1437H}.
 The sample contains stars with spectral types between K7V and M4.5V.
 These stars are listed in Table~\ref{sptype_calibration_sample}.


\begin{table}
\centering
\caption{Spectral Type calibration sample.}
\label{sptype_calibration_sample}
\begin{tabular}{lll|lll}
\hline\noalign{\smallskip}
 GJ     &       SpT     &   Ref$^{\dag}$ &  GJ  &       SpT     &   Ref$^{\dag}$ \\
\hline
185   &       K7V     &       h & 408   &       M2V     &       h \\
686   &       M0V     &       h & 250B  &       M2.5V   &       k \\
701   &       M0V     &       h & 581   &       M2.5V   &       h \\
846   &       M0.5V   &       k & 352A  &       M3V     &       k \\
720A  &       M0.5V   &       k & 436   &       M3V     &       k \\
229   &       M1V     &       k & 752A  &       M3V     &       k \\
412A  &       M1V     &       h & 569A  &       M3V     &       k \\
514   &       M1V     &       h & 273   &       M3.5V   &       k \\
570B  &       M1V     &       h & 643   &       M3.5V   &       k \\
908   &       M1V     &       h & 734B  &       M3.5V   &       k \\
205   &       M1.5V   &       k & 213   &       M4V     &       k \\
15A   &       M1.5V   &       h & 402   &       M4V     &       k \\
526   &       M1.5V   &       h & 699   &       M4V     &       k \\
625   &       M1.5V   &       h & 83.1  &       M4.5    &       k \\
220   &       M2V     &       k & 166C  &       M4.5    &       k \\
382   &       M2V     &       k & 234A  &       M4.5    &       k \\
393   &       M2V     &       h &       &               &         \\
\hline
\end{tabular}
\tablefoot{$^{\dag}$ 
k: standard from \cite{1991ApJS...77..417K}; 
h: standard from \cite{1994AJ....108.1437H}.}
\end{table}

 Spectral type-sensitive ratios were identified by means of a
 Spearman's correlation test.  
 For each of them a third order polynomial
 fit was performed between the numerical spectral-type index (with value 0.0 for
 M0; 0.5 for M0.5 and so on) and the ratio. A negative value implies
 that the star is a late-K dwarf instead of an M star
 being the index value for K7 equal to -1.0
 (K7 is the subtype preceding M0).  
 
\begin{figure}  
\centering
\includegraphics[angle=270,scale=0.55]{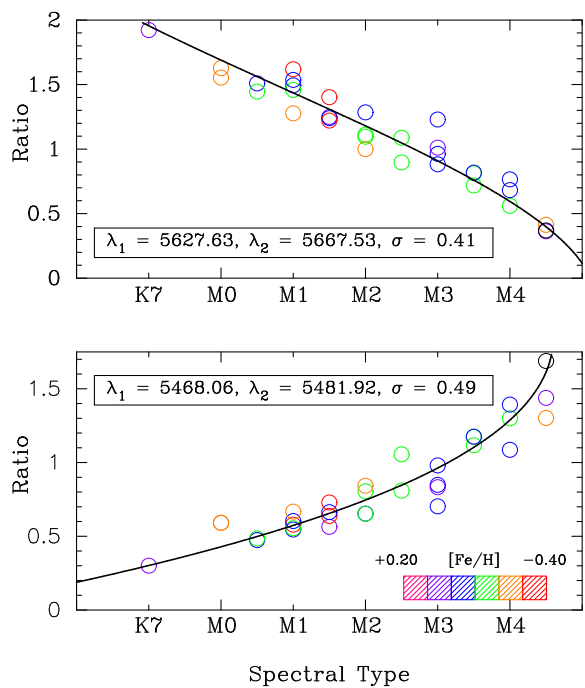}
\caption{
Spectral type as a function of two different spectral-type sensitive ratios.
Stars are plotted using different colours according to their metallicity
(using 0.10 dex length bins).
 A third order polynomial fit is shown. The features' central wavelengths as well as the rms standard
deviation on spectral type of the residuals are given in each plot.
}
\label{examples_of_sptype_ratios}
\end{figure}

 The third order polynomial
 fit was preferred amongst other functional fits since we found it to give
 the lowest rms standard deviation. 
 Our ``final'' selection of spectral type-sensitive ratios includes 82
 ratios of features with a standard deviation lower than 0.5 spectral subtypes. 
 The derived mean numerical spectral types are rounded to the nearest half integer. 
 A couple of examples are shown in Figure~\ref{examples_of_sptype_ratios}.
 Full details for some examples are given in Table~\ref{calibrations_sptype}.
 As for T$_{\rm eff}$, we performed 1000 simulations with random
 spectral-type values. The results show that we are not able to find
 any suitable calibration in any of the simulations when random spectral-types are 
 used.

\begin{table}
\centering
\caption{
Coefficients of our spectral type calibrations. Columns (1) and (2) show the wavelength of the
corresponding features, columns (3) to (6) the coefficients of the fit,
while column (7) gives the standard deviation of the calibration of spectral types.
Only four examples are shown here.} 
\label{calibrations_sptype}
\begin{tabular}{ccccccc}
\hline\noalign{\smallskip}
$\lambda_{\rm 1}$ &  $\lambda_{\rm 2}$ & a$_{0}$  & a$_{1}$  & a$_{2}$  & a$_{3}$ & $\sigma$ \\
(1) & (2) & (3) & (4) & (5) & (6) & (7) \\
\hline
5468.06 & 5481.92 & -3.88 & 10.79 & -4.25 & 0.48 & 0.49 \\
5627.63 & 5667.53 & 5.11  & -0.75 & -2.18 & 0.50 & 0.41 \\
5467.06 & 5481.92 & -3.80 & 8.18  & -0.56 & -0.77& 0.48 \\
5467.06 & 5512.54 & -5.09 & 10.67 & -4.13 & 0.53 & 0.45 \\
\hline
\end{tabular}
\end{table}

\subsection{Metallicity}
\label{metallicity}

 A common approach to find metallicity calibrators for low-mass stars relies
 on the search of M dwarfs in common proper motion pairs orbiting around
 a solar-type star with accurate spectroscopic metallicity determinations.
 The basic assumption is that both stars are coeval and born in the same
 protostellar cloud so  the 
 metallicity of the secondary M dwarf is the same as the one of the primary
 star \citep[e.g.][]{2005A&A...442..635B}. 
 Until very recently, only the most nearby and bright M stars
 have been searched for planets by means of the Doppler technique.
 As a consequence there is a lack of HARPS and HARPS-N spectra for most of the identified
 M dwarfs in binary systems around solar-type stars.
 To overcome this difficulty we built a list of 47 metallicity calibrators with available HARPS spectra,
 known parallaxes, and magnitudes by using the most
 recent photometric calibration provided in NE12. 
 This calibration is in turn based on metallicity determinations from FGK
 primaries with an M dwarf secondary. 
 The sample of metallicity calibrators covers a wide range in metallicity from  -0.54 to +0.24 dex
 with
 typical error bars of the order of
 $\lesssim$ 0.05 dex. These stars are listed in Table~\ref{metallicity_calibration_sample}.
  We caution that the uncertainties
 reported in Table~\ref{metallicity_calibration_sample}
 do not take into account the scatter in the NE12 calibration, which is
 of the order of $\sim$ 0.17 dex.

\begin{figure}[!htb]
\centering
\includegraphics[angle=270,scale=0.45]{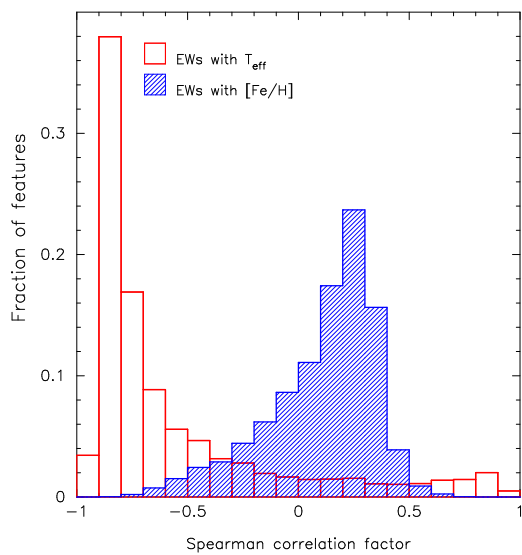}
\caption{
Spearman's correlation factor distribution
of the EW with the stellar metallicity (blue histogram) and
with the T$_{\rm eff}$ (red histogram). The sample of metallicity calibrators
and the initial list of 4224 features is considered. 
Metallicities are computed using NE12 whilst effective temperatures
are obtained with our methodology.}
\label{histograma_partial}
\end{figure}

 The effects of metallicity on the EW of the features are 
 entangled with the effects of T$_{\rm eff}$,
  with T$_{\rm eff}$ as the primary driver of changes 
 in the EW.
 This can be easily seen in the histograms
 in Figure~\ref{histograma_partial}. They show the distribution of the Spearman's rank correlation factor
 of the EW with the stellar metallicity and with
 T$_{\rm eff}$.
 The figure shows that a significant fraction of the features,  $\sim$ 43\%, shows a high
 (Spearman's correlation factor $<$ -0.80)
 anticorrelation with T$_{\rm eff}$, while only a relatively small fraction
  ($\sim$ 3\%) shows a significant positive correlation.
 The correlation between stellar metallicity and EWs is generally less
 significant. 
 The distribution of Spearman's coefficients for metallicity  
 shows a clear peak at +0.25 which drops almost to zero at +0.50, while at negative values it has
 a smooth  tail down to -0.80.  
 Effects of metallicity and effective temperature
 should therefore be considered simultaneously.

 We 
 searched for empirical relationships  for metallicity as a function
 of features and ratios of features  (i.e., an indicator of temperature) 
 with the following analytical form:  

 \begin{equation}
 [Fe/H] = (A\times EW) + (B\times r) + C
 \end{equation}

 \noindent where $EW$ is the EW of a feature
 showing a strong-metallicity correlation, $r$ is a temperature-sensitive
 ratio of features, and $A$, $B$, $C$ are independent coefficients.
 We considered every combination of features and ratios satisfying the condition that
 the correlation of $EW$ with metallicity, and the correlation of $r$ with T$_{\rm eff}$ show at least a 98\%
 of significance. 
 Our final selection consists of  696 calibrations with standard
 deviation values between 0.07 and $\sim$ 0.10 dex.
  We point out that these uncertainties should be considered as
 lower limits since 
 they do not take into account possible
 systematic errors in the underlying NE12 calibration.

  As before, we performed a series of simulations
 using random metallicities and errors. In 84\% of the simulations we do not find
 any suitable metallicity calibration, although in 7.5\% of the cases the simulation
 finds a large number of calibrations (larger than 348).
 Some examples  of our obtained metallicity calibrations are
 provided in Table~\ref{calibrations_metallicity}.

\begin{table}
\centering
\caption{Metallicity calibration sample. [Fe/H] are computed
from V magnitudes and parallaxes taken from the compilation of \cite{2013A&A...549A.109B} 
and 2MASS magnitudes \citep{2003yCat.2246....0C}. The photometric calibration by NE12 is used. 
Errors are computed by propagating the uncertainties in the parallaxes and the
photometry.
}
\label{metallicity_calibration_sample}
\begin{tabular}{lr|lr}
\hline\noalign{\smallskip}
 Star   &       [Fe/H]                  &  Star         &       [Fe/H]                  \\
\hline
GJ1	&	-0.37	$\pm$	0.04	&	GJ551	&	0.13	$\pm$	0.04	\\
GJ105B	&	-0.13	$\pm$	0.03	&	GJ555	&	0.10	$\pm$	0.04	\\
GJ176	&	0.02	$\pm$	0.04	&	GJ569A	&	0.02	$\pm$	0.03	\\
GJ205	&	-0.03	$\pm$	0.19	&	GJ588	&	0.05	$\pm$	0.03	\\
GJ2066	&	-0.10	$\pm$	0.03	&	GJ618A	&	-0.06	$\pm$	0.04	\\
GJ213	&	-0.24	$\pm$	0.04	&	GJ628	&	-0.05	$\pm$	0.03	\\
GJ229	&	-0.02	$\pm$	0.17	&	GJ674	&	-0.20	$\pm$	0.03	\\
GJ250B	&	-0.10	$\pm$	0.04	&	GJ678.1A&	-0.14	$\pm$	0.04	\\
GJ273	&	-0.11	$\pm$	0.03	&	GJ680	&	-0.05	$\pm$	0.04	\\
GJ300	&	0.06	$\pm$	0.03	&	GJ682	&	0.10	$\pm$	0.03	\\
GJ357	&	-0.32	$\pm$	0.03	&	GJ686	&	-0.30	$\pm$	0.03	\\
GJ358	&	0.05	$\pm$	0.03	&	GJ693	&	-0.29	$\pm$	0.03	\\
GJ367	&	-0.05	$\pm$	0.04	&	GJ701	&	-0.19	$\pm$	0.03	\\
GJ382	&	0.05	$\pm$	0.03	&	GJ752A	&	0.01	$\pm$	0.03	\\
GJ393	&	-0.11	$\pm$	0.04	&	GJ832	&	-0.17	$\pm$	0.04	\\
GJ413.1	&	-0.11	$\pm$	0.04	&	GJ846	&	-0.07	$\pm$	0.04	\\
GJ438	&	-0.51	$\pm$	0.07	&	GJ849	&	0.24	$\pm$	0.04	\\
GJ447	&	-0.26	$\pm$	0.03	&	GJ876	&	0.13	$\pm$	0.03	\\
GJ465	&	-0.54	$\pm$	0.04	&	GJ877	&	-0.02	$\pm$	0.03	\\
GJ479	&	0.05	$\pm$	0.04	&	GJ887	&	-0.35	$\pm$	0.14	\\
GJ514	&	-0.11	$\pm$	0.04	&	GJ908	&	-0.39	$\pm$	0.03	\\
GJ526	&	-0.16	$\pm$	0.03	&	HIP31293&	0.03	$\pm$	0.04	\\
GJ536	&	-0.15	$\pm$	0.04	&	LTT9759	&	0.14	$\pm$	0.04	\\
GJ54.1	&	-0.47	$\pm$	0.05	&		&				\\
\hline
\end{tabular}
\end{table}

\begin{table}
\centering
\caption{
Coefficients of our metallicity calibrations. Column (1) shows the wavelength of the
corresponding feature, column (2) the temperature-sensitive ratio, columns (3) to (5)
the coefficients $A$, $B$, and $C$, 
while column (6) gives the standard deviation of the calibration.  
Only five examples are shown here.}
\label{calibrations_metallicity}
\begin{scriptsize}
\begin{tabular}{cccccc}
\hline\noalign{\smallskip}
$\lambda_{1}$ & $\lambda_{2}$/ $\lambda_{3}$ & $A$ & $B$ & $C$ & $\sigma$ (dex) \\
(1) & (2) & (3) & (4) & (5) & (6)  \\ 
\hline
6785.77	&	6785.38/6799.25	&	-0.0354	&	-1.876	&	1.753	&	0.07	\\
6785.77	&	6799.25/6785.38	&	-0.0349	&	0.227	&	0.416	&	0.07	\\
6785.77	&	6785.38/6790.93	&	-0.0359	&	-0.930	&	1.633	&	0.07	\\
6785.77	&	6785.38/6788.76	&	-0.0375	&	-1.163	&	1.581	&	0.07	\\
6785.77	&	6790.93/6785.38	&	-0.0356	&	0.263	&	0.613	&	0.07	\\
\hline
\end{tabular}
\end{scriptsize}
\end{table}

\section{Comparison with other methods}
\label{comparison}

 Our calibrations have been applied to a sample of 53 M dwarfs
 from the HARPS GTO M dwarf sample \citep{2013A&A...549A.109B}
 for which HARPS data have been obtained from the ESO Science 
 Data Products Archive\footnote{archive.eso.org/wdb/wdb/adp/phase3\_spectral/form?phase3\_collection\\=HARPS}.
 Only spectra with a median signal-to-noise ratio of at least 25 were considered. 
 For stars with more than one spectrum available we took the one with the highest signal-to-noise ratio.
 No further restrictions were applied. 
 The sample is composed of nearby (distance $<$ 11 pc),
 bright  (V $<$ 12, K$_{\rm S}$ $<$ 7), early-type M dwarfs
 (spectral types M0V-M4.5V). 
 Our methods were applied to compute effective temperatures,
 stellar metallicities and to derive spectral-types. ``Final'' values
 for these parameters are the mean of the individual values from
 all the calibrations.   
 All these quantities are provided in Table~\ref{parameters_table_full},
 which is available in the online version of this paper. 

 Our results are compared with:
 {\it i)} A photometric scale, namely CA08 for T$_{\rm eff}$ and
  NE12 for [Fe/H];
 {\it ii)} The recent work by \citet[][hereafter GA14]{2014MNRAS.443.2561G};
 and {\it iii)} The values obtained with the methodology
 developed by NE14. 

\subsection{Comparison of effective temperatures}
\label{subseccomparteff}

\begin{figure}[!htb]
\centering
\includegraphics[angle=270,scale=0.50]{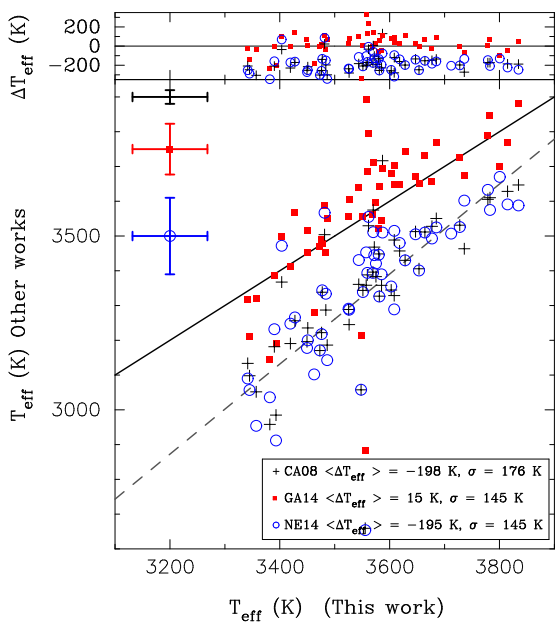}
\caption{
T$_{\rm eff}$ values from the literature estimates versus the values obtained in this work. 
The upper
panel shows the differences between the temperatures 
given in the literature and the values derived in this work.
Median uncertainties in the derived temperatures are also shown.
The symbol $<>$ in the legend represents the median difference. 
The black continuous line represents the 1:1 relation whilst
the grey dashed one
represents the best linear fit between our estimates and those obtained using
CA08 relationship (see text in Section~\ref{subseccomparteff}).}
\label{comparison_temperatures}
\end{figure}

 Photometric effective temperatures are derived from
 V magnitudes from the compilation of \cite{2013A&A...549A.109B}
 and 2MASS  \citep{2003yCat.2246....0C}  photometry using
 the calibration provided by CA08. 
 Computed errors take into account the propagation of the uncertainties
 of the 2MASS magnitudes as well as the accuracy of the CA08 calibrations.
 The comparison between the photometrically derived temperatures and our
 spectroscopic ones is illustrated in Figure~\ref{comparison_temperatures}.
 There is a clear offset between our spectroscopic estimates and the 
 photometric temperatures, being the latter 
 cooler than ours (the median
 difference $\Delta$T$_{\rm eff}$ = T$^{\rm phot}_{\rm eff}$ - T$^{\rm spec}_{\rm eff}$ 
 is  -198 K with a rms standard deviation of  176 K).
 Our temperatures can be converted into the CA08 scale by a 
 linear transformation:
 T$_{\rm eff}$[CA08 scale] = (1.29 $\pm$ 0.02)$\times$T$_{\rm eff}$[this work]
 + (-1271 $\pm$ 89) K (dashed grey line).

\begin{figure}[!htb]
\centering
\includegraphics[angle=270,scale=0.50]{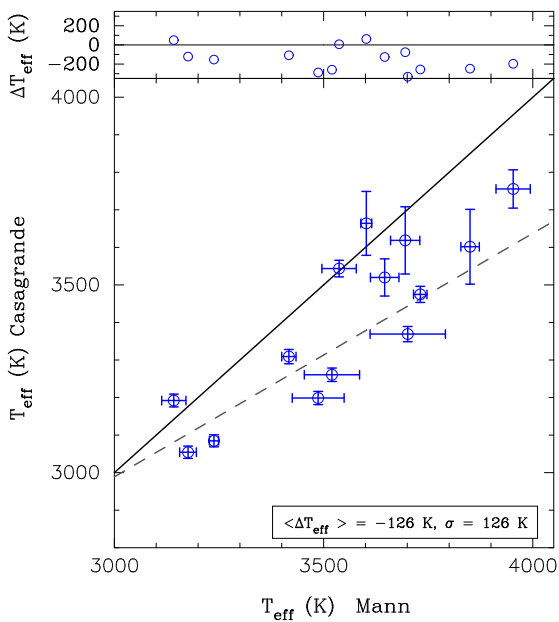}
\caption{
T$_{\rm eff}$ estimates based on
the calibration provided by CA08,
versus values from \cite{2013ApJ...779..188M} and \cite{2014arXiv1412.2758N}.
The upper
panel shows the differences.
The symbol $<>$ in the legend represents the median difference.
The black continuous line represents the 1:1 relation whilst
the grey dashed one
represents the best linear fit.}
\label{comparison_boya_casagrande}
\end{figure}

 The reason for this discrepancy is  
 the choice  of \cite{2013ApJ...779..188M} temperatures as our reference temperature scale. 
 In order to test this, photometric T$_{\rm eff}$ values were computed for our
 sample of temperature calibrators (Table~\ref{calibration_sample}) following CA08.
 The comparison between  \cite{2013ApJ...779..188M} and CA08 temperatures is shown
 in Figure~\ref{comparison_boya_casagrande}. 
 It can be seen from this figure that CA08 temperatures tend to be systematically
 lower than those provided by  \cite{2013ApJ...779..188M}.  

 The discrepancy between CA08 values and interferometry-based temperatures
 has also been noted in a recent work by \cite{2015arXiv150101635M}.
 The difference $\Delta$T$_{\rm eff}$ between CA08 and interferometric-based
 temperatures noted by these authors is 160 K (the CA08 temperatures being
 cooler) in agreement with our results. 
 CA08 temperatures are obtained by extending the IRFM
 for FGK dwarfs from \cite{2006MNRAS.373...13C}  to M dwarfs.
 One of the assumptions of the IRFM is that a star can be 
 approximated as a blackbody for wavelengths beyond $\approx$  2 $\mu$m.
 \cite{2015arXiv150101635M} argue that whilst this assumption is reasonable
 for warmer stars, it does not suit M dwarfs, which have significantly
 more flux in the near-infrared than predicted by a blackbody. 
 As a consequence CA08 temperatures tend to be systematically lower, with increasing
 disparity at cooler temperatures where stars deviate more from
 the blackbody emission (see Figures~\ref{comparison_temperatures} and
 \ref{comparison_boya_casagrande}).
 \cite{2015arXiv150101635M} also note that the temperature scale of the old
 version of the PHOENIX models used in CA08 
 differs from interferometric-based temperatures. 

 We have also compared our temperatures with the values 
 given by NE14.
 Since NE14
 method is calibrated using the CA08 photometric relationship, 
 the comparison of our temperatures with NE14 shows results similar to the comparison
 with CA08 (Figure~\ref{comparison_temperatures}).

 Our sample contains 51 stars in common with the sample of GA14 
 who determine effective temperatures by comparing low-medium resolution
 spectra with the BT-SETTL version of the PHOENIX model atmospheres
 \citep{2012RSPTA.370.2765A,2012EAS....57....3A}.
 Their procedure was calibrated using the
 stars listed in BO12, although with the stellar bolometric fluxes
 computed as in \cite{2013ApJ...779..188M}.
 As can be seen in Figure~\ref{comparison_temperatures}, GA14 temperatures
 tend to be slightly hotter than ours, specially for T$_{\rm eff}$ $>$ 3400 K.
 For the coolest dwarfs in this sample, GA14 temperatures depart from ours and tend
 to be smaller than ours. 

\subsection{Comparison of metallicities}
\label{subseccomparmetal}

\begin{figure}[!htb]
\centering
\includegraphics[angle=270,scale=0.50]{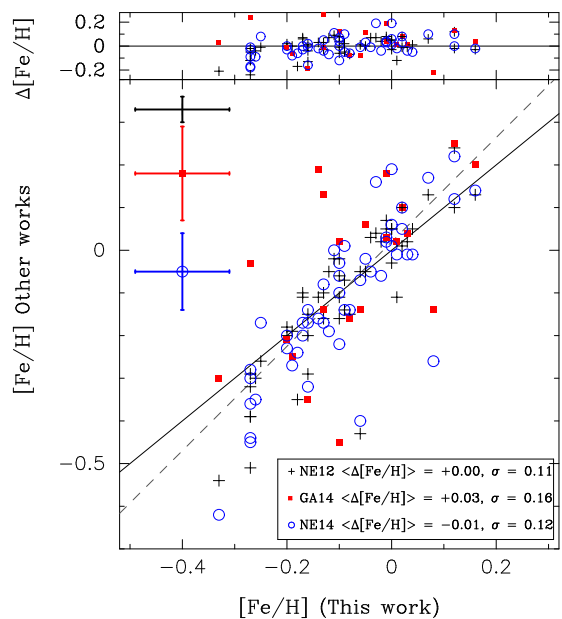}
\caption{
[Fe/H]  values from the literature estimates versus the values obtained in this work.
The upper
panel shows the differences between the metallicities
given in the literature and those derived in this work.
Median uncertainties in the derived metallicities are also shown.
The symbol $<>$ in the legend represents the median difference.
The black continuous line represents the 1:1 relation whilst
the grey dashed one
represents the best linear fit between our estimates and those obtained using
NE12 relationship  (see text in Section~\ref{subseccomparmetal}).}
\label{comparison_metallicity}
\end{figure}

 We also compare our metallicities with those reported  previously in the literature.
 Values for the comparison are taken
 from the photometric calibration by NE12; 
 from  GA14 
 who determine metallicities
 following the method of \cite{2013ApJ...779..188M} based on empirical calibrations 
 between the strength of atomic and molecular spectroscopic features and
 stellar metallicity; and from NE14. 
 The comparison is shown in Figure~\ref{comparison_metallicity}.
 
 The comparison reveals an overall good agreement between our
 metallicity estimates and those by NE12, GA14, and NE14.
 The median differences with these works are consistent with
 zero and the scatter although somewhat large is consistent
 within the (also large) error bars. 
 A linear fit between our metallicities and those obtained
 using NE12 relationship provides a slope slightly larger than 
 one and a small difference in the zero point:
 [Fe/H][NE12 scale] = (1.22 $\pm$ 0.04)$\times$[Fe/H][this work]
 + (0.02 $\pm$ 0.01) dex (dashed grey line). 

\subsection{Comparison of spectral types}

\begin{figure}[!htb]
\centering
\includegraphics[angle=270,scale=0.50]{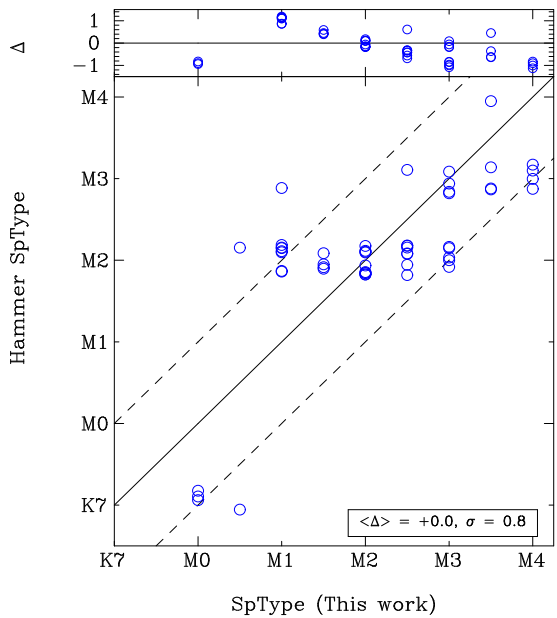}
\caption{
M subtype  values  obtained by using the HAMMER code
 versus the values obtained in this work.
 The upper
panel shows the differences with 
the values given in the literature.
Random values
between $\pm$0.2 have been added to the 
Hammer values to help in the comparison.
The symbol $<>$ in the legend represents the median difference.
The black continuous line represents the 1:1 relation whilst
the dashed ones correspond to $\pm$ 1 spectral subtype. }
\label{comparison_sptypes}
\end{figure}

 We finally compare the spectral types derived by us
 with those obtained by using
 the automatic procedure of the HAMMER spectral code
 \citep[][hereafter CO07]{2007AJ....134.2398C}.
 The CO07 code was designed to classify stars in the Sloan
 Digital Sky Survey Spectroscopic database, therefore before using it
 our spectra were degraded to a resolution R $\sim$
 2000 by convolving them with a gaussian profile. 
 We also caution that roughly half of the spectral-type sensitive
 band indices defined in CO07 are outside the HARPS
 spectral coverage. 
 The comparison is shown in Figure~\ref{comparison_sptypes}.
 It can be seen that there seems to be no significant differences
 between our estimates and those by CO07, with an
 overall good agreement 
 within $\pm$ 1 spectral subtype (dashed lines in
 Figure~\ref{comparison_sptypes}). 
 \cite{2011AJ....141...97W} and \cite{2013AJ....145..102L} found the automatic spectral types given by the Hammer
 code
 to be about one subtype earlier than the manual classification.
 While this effect is not evident in our comparison it can not
 be ruled out either. 


\begin{figure}[!htb]
\centering
\includegraphics[angle=270,scale=0.50]{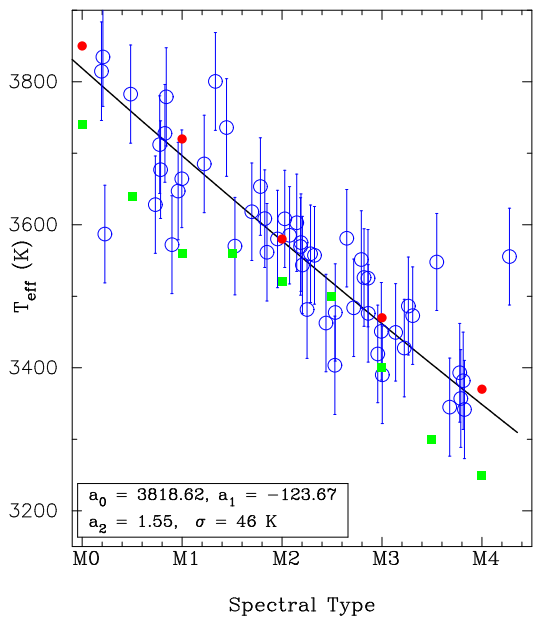}
\caption{
 Effective temperature as a function of the spectral type. 
 For clarity
 the spectral types are not rounded.
 A second order polynomial fit is shown.
 Possible outliers are removed by using a 2.5$\sigma$ clipping
 procedure. 
 The coefficients of the fit
 as well as the rms standard deviation are given in the plot. 
 Data from \citet[][Table~A.5]{1995ApJS..101..117K}
 is overplotted using red filled circles, whilst green squares
 represent the median T$_{\rm eff}$-spectral type sequence from
 \cite{2013AJ....145..102L}.}
\label{sptype_teff}
\end{figure}

  Figure~\ref{sptype_teff} shows our derived effective temperatures as a function
  of the spectral type. For comparison data from \citet[][Table~A.5]{1995ApJS..101..117K}
  is overplotted (red circles).  The median T$_{\rm eff}$-spectral type sequence from
  \citet{2013AJ....145..102L} is also shown (green squares).
  It can be seen that except for the presence of some outliers,
  effective temperatures and spectral types are well correlated following
  the expected trend. In other words, our spectral types appear  
  to be consistent with our temperature scale. Unlike  \cite{2013AJ....145..102L},
  our data do not suggest the presence of a T$_{\rm eff}$ plateau 
  in the spectral range M1-M3 although we note that our sample is relatively
  small in comparison with the one in \cite{2013AJ....145..102L}.
 
 We conclude that 
 our metallicities and spectral types agree reasonably well with other
 literature estimates. 
 Regarding effective temperatures, 
 there is a clear offset between BO12 and CA08 scales, as explained.
 In summary, our methodology can be confidently used to characterise
 large samples of stars in an homogeneous way. 

\section{Empirical relationships for stellar mass, radius, and gravity}
\label{mass_radii_gravity}

 We made use of the temperature and metallicity values derived with our
 method to search for 
 empirical relationships with the stellar evolutionary parameters
 namely, stellar mass, radius, and surface gravity.

\begin{figure}[!htb]
\centering
\includegraphics[angle=270,scale=0.50]{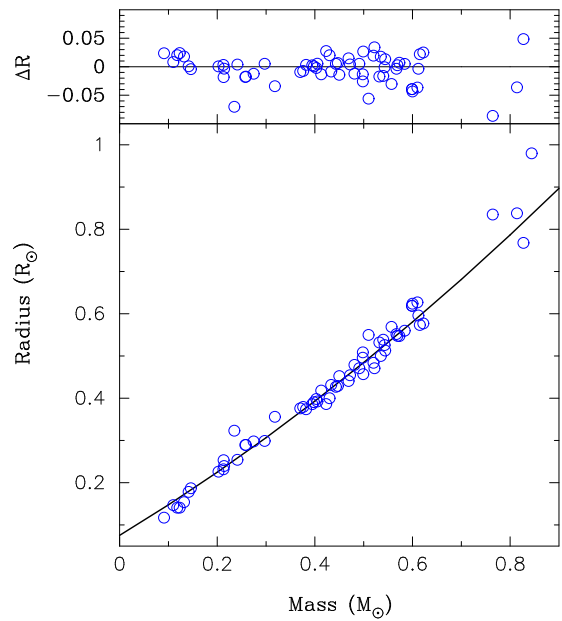}
\caption{
 Stellar radius as a function of the stellar mass. The sample includes 
 stars with interferometric measurements of their radii from
 BO12 and \cite{2014MNRAS.438.2413V} as well as low-mass eclipsing
 binaries from \citet[][Table~5]{2014arXiv1408.1758H}. 
 The best fit is also shown. A  3$\sigma$ clipping procedure was
 used to remove outliers. The upper panel shows the differences
 between the radius derived with our fit and the radius given
 in the literature. Median errors in radius (not shown in the plot)
 are of the order of 0.006 R$_{\odot}$.} 
\label{mass_radio}
\end{figure}

  We derived our own mass-radius relationship by combining the
  stars with known interferometric radius from BO12 and \cite{2014MNRAS.438.2413V} 
  with data from low-mass eclipsing binaries provided in the compilation
  by \citet[][Table~5]{2014arXiv1408.1758H}. A 3$\sigma$ clipping procedure was
  used to remove potential outliers. Our derived calibration is as 
  follows:

  \begin{equation} 
  \label{eq_masa_radio}
    R = 0.0753 + 0.7009\times M  + 0.2356 \times M^2 
  \end{equation}

  \noindent where radius and masses are given in solar units and the rms
   standard deviation of the calibration is 0.02 R$_{\odot}$. 
   The radius-mass plane is shown in Figure~\ref{mass_radio}.

 Values of stellar
 masses were obtained for each of our target stars
 following the relations based on near infrared
 photometry by \cite{1993AJ....106..773H}.
 We chose this calibration since it is the same used 
 by BO12. 
 These calibrations are provided in the CIT photometric system therefore,
 before applying them, 2MASS magnitudes
 were converted into CIT magnitudes following the transformations
 provided by \cite{2001AJ....121.2851C}.
 Once the stellar masses were computed, values of the radius
 were derived using Equation~\ref{eq_masa_radio}. Finally,
 surface gravities, $\log g$, were derived from masses and
 radii. 

\begin{figure*}[!htb]
\centering
\begin{minipage}{0.47\linewidth}
\includegraphics[angle=270,scale=0.45]{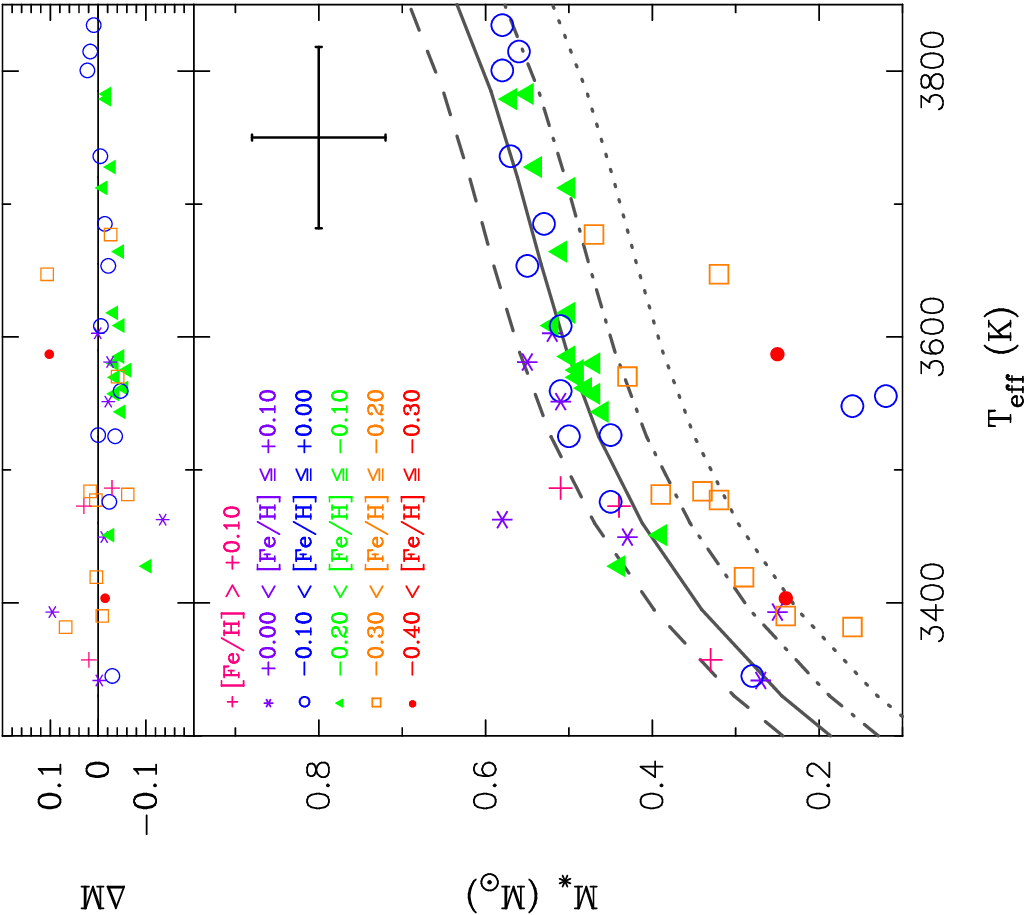}
\end{minipage}
\begin{minipage}{0.47\linewidth}
\includegraphics[angle=270,scale=0.45]{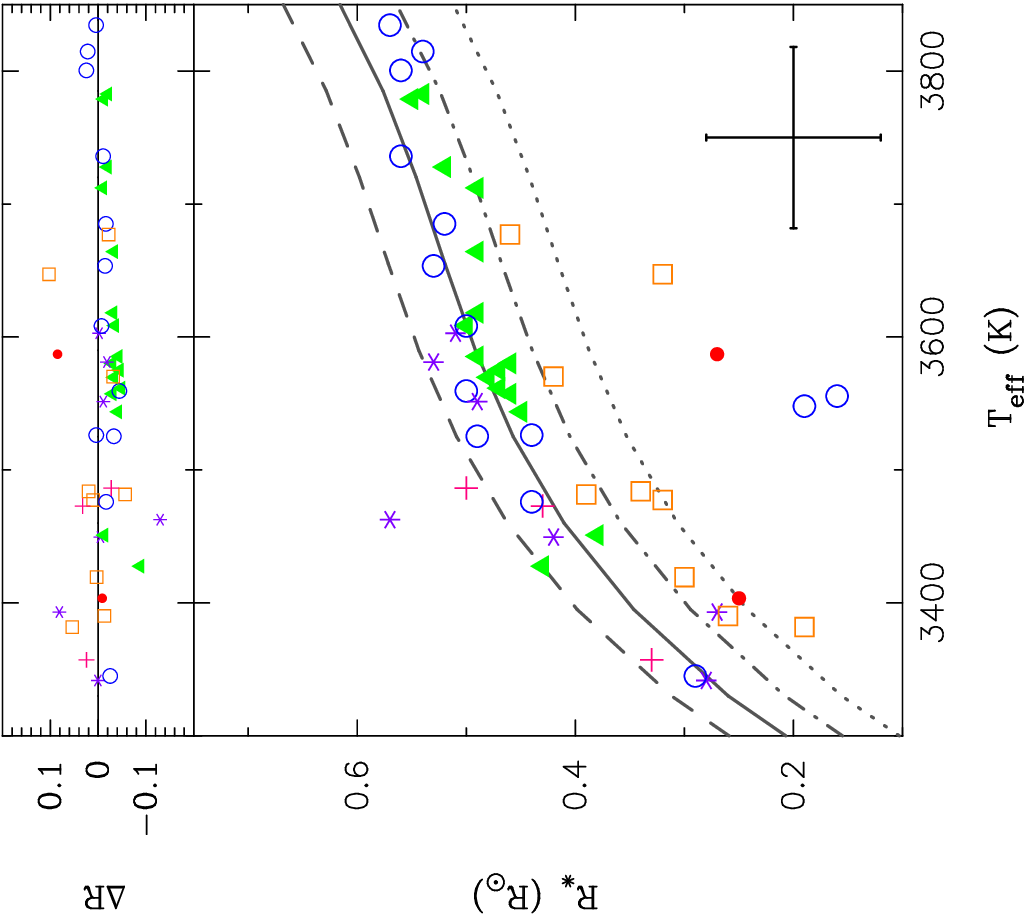}
\end{minipage}
\caption{
 Stellar mass (left panel), and radius (right panel), 
 as a function of the effective temperature. Stars are plotted
 using different colours and symbols according to their metallicity. Several fits for fixed
 metallicity values are plotted: +0.15 (dashed line), +0.00 (solid line),
 -0.15 (dash-dotted line), and -0.30 (dotted line).
 The upper left panel shows the differences between the mass
 obtained from Equation~\ref{eqn:m} and those derived by using
 \cite{1993AJ....106..773H} relationship.
 The upper right panel shows the differences between the radius
 derived from Equation~\ref{eqn:r} and by using
 Equation~\ref{eq_masa_radio}.}
\label{mrg_teff_metal_plot}
\end{figure*}

 We first investigated the correlation of M$_{\star}$, R$_{\star}$, and
 $\log g$ with the effective temperature and with the stellar metallicity by using the Spearman
 correlation test. Results are given in Table~\ref{mrg_teff_metal}.
 It can be seen that although the main dependence of the evolutionary parameters
 is on the effective temperature, they also show a moderate but significant
 dependence on the stellar metallicity. 

 We also evaluated the significance of the correlations by a bootstrapp
 Monte Carlo (MC) test plus a gaussian, random shift of each
 data-point within its error bars. For each pair of variables 10000 random
 datasets were created, determining the coefficient of
 correlation, $\rho$, and its corresponding z-score each time.
 The tests were done using the code 
 {\sc MCSpearman}\footnote{https://github.com/PACurran/MCSpearman/} 
 by \citet{2014arXiv1411.3816C} which might be consulted for
 further details on this method. 
 Mean values 
 of $\rho$ and z-score are
 given in Table~\ref{mrg_teff_metal}. 
 We note that the results from the MC method do not exclude a metallicity dependence but suggest
 that any possible correlation is relatively weak. 
 
\begin{table}
\centering
\caption{Results from the Spearman's correlation test and the MC method showing the dependence of the
 evolutionary parameters on the effective temperature, and on the stellar metallicity.}
\label{mrg_teff_metal}
\begin{tabular}{lcccc}
\hline\noalign{\smallskip}
             & \multicolumn{4}{c}{T$_{\rm eff}$}                                     \\
             & \multicolumn{2}{c}{Spearman's test}  & \multicolumn{2}{c}{MC method}  \\
             & \multicolumn{2}{c}{\hrulefill}       & \multicolumn{2}{c}{\hrulefill} \\ 
             & $\rho$    &  prob.                   & $\rho$ & z-score               \\
\hline
 Mass        &  0.72     &  2*10$^{\rm -9}$        & 0.47  &  1.55                  \\
 Radius      &  0.72     &  2*10$^{\rm -9}$        & 0.47  &  1.55                  \\   
 $\log g$    & -0.73     &  2*10$^{\rm -9}$        & -0.30 & -0.93                  \\
\hline
             & \multicolumn{4}{c}{[Fe/H]}                                     \\
             & \multicolumn{2}{c}{Spearman's test}  & \multicolumn{2}{c}{MC method}  \\   
             & \multicolumn{2}{c}{\hrulefill}       & \multicolumn{2}{c}{\hrulefill} \\      
             & $\rho$    &  prob.                   & $\rho$ & z-score               \\
\hline
 Mass        &  0.34   &  0.02 & 0.20  & 0.60  \\
 Radius      &  0.34   &  0.02 & 0.20  & 0.60  \\
 $\log g$    & -0.33   &  0.02 & -0.12 & -0.40 \\
\hline
\end{tabular}
\end{table}

 A dependence of the radius on stellar metallicity
 is expected from model predictions
 \citep{1998A&A...337..403B,2008ApJS..178...89D} however
 BO12  find the interferometry-based  radius rather insensitive to metallicity.
 Furthermore, while fitting
 M$_{\star}$, R$_{\star}$, and L$_{\star}$ as a function
 of the effective temperature \cite{2013ApJ...779..188M}
 find that adding the stellar
 metallicity as a parameter does not improve the fits.
 However, in a more recent work, \cite{2015arXiv150101635M}
 do find a significant effect of the metallicity on 
 the T$_{\rm eff}$-R$_{\star}$ relation. The authors
 point towards small sample sizes and a sparser sampling
 on [Fe/H] as the reasons why the effect of [Fe/H]
 was not noticed in their previous studies. 
 We therefore performed two kinds of fit, one using only
 the effective temperature and another one adding the 
 stellar metallicity as a parameter. 
 The extra sum-of-squares F test \citep[e.g.][]{1993stp..book.....L}
 was used to test whether the addition of the metallicity
 to the functional form of the calibrations provides
 any improvement or not. 
 The test returns  
 a measure of the likelihood ($p$-value)
 that the simpler model (the one with fewer parameters) 
 provides a better representation than the more complicated one. 
 The resulting values\footnote{The tests were performed using
 the {\sc MPFTEST} IDL routine
 included in the {\sc MPFIT} package \citep[][]{2009ASPC..411..251M}
 and available at  http://cow.physics.wisc.edu/~craigm/idl/idl.html},
 $p$-M$_{\star}$ $\sim$ 2$\times$10$^{\rm -7}$,  
 $p$-R$_{\star}$ $\sim$ 7$\times$10$^{\rm -8}$, 
 $p$-$\log g$ $\sim$ 2$\times$10$^{\rm -7}$,  
 indicate that by including the metallicity there is improvement in the fits
 in line with \cite{2015arXiv150101635M}.
 The relationships we obtain are the following: 

\begin{eqnarray}
\label{eqn:m} \nonumber M_{\star} (M_{\odot}) = -171.616 + 0.139\times T_{\rm eff} - 3.776\times10^{\rm -5}T_{\rm eff}^{\rm 2}\\+ 3.419\times10^{\rm -9}T_{\rm eff}^{\rm 3} + 0.382 \times [Fe/H] \\
\label{eqn:r} \nonumber R_{\star} (R_{\odot}) = -159.857 + 0.130\times T_{\rm eff} - 3.534\times10^{\rm -5}T_{\rm eff}^{\rm 2}\\+ 3.208\times10^{\rm -9}T_{\rm eff}^{\rm 3} + 0.347 \times [Fe/H] \\
\label{eqn:g} \nonumber \log g (cgs) = 174.462 -0.138\times T_{\rm eff} + 3.728\times10^{\rm -5}T_{\rm eff}^{\rm 2}\\- 3.376\times10^{\rm -9}T_{\rm eff}^{\rm 3} - 0.332 \times [Fe/H] 
\end{eqnarray}

 \noindent where the rms standard deviations of the calibrations are $\sigma_{\rm M\star}$ = 0.02 M$_{\odot}$,
 $\sigma_{\rm R\star}$ = 0.02 R$_{\odot}$ , and $\sigma_{\log g}$ = 0.02 (cgs). 
 The calibrations are valid for 3340 K $<$ T$_{\rm eff}$ $<$ 3840 K,
 and -0.40 $<$ [Fe/H] $<$ +0.16 dex.
 Empirical relationships for stellar luminosity are not provided since they
 can be easily obtained from T$_{\rm eff}$ and R$_{\star}$ just applying
 Stefan-Boltzmann's law. 
 All these quantities (M$_{\star}$,  R$_{\star}$, $\log g$, and $\log (L_{\star}/L_{\odot})$ )
 for the stars analyzed in this work are provided in Table~\ref{parameters_table_full}.

 Typical uncertainties are in the order of 13.1\% for the stellar mass,
 11.8\% for the radius, 25\% for
 luminosities, and 0.05 dex for $\log g$.
 We note that these uncertainties are computed by taking into account the 
 $\sigma$ of the corresponding calibration and the propagation of the 
 errors in T$_{\rm eff}$ and [Fe/H].
 A word of caution should be given regarding the relative errors in mass
 since they tend to increase towards lower masses. 
 Relative errors in mass might be larger than 20\% for stars with
 M$_{\star}$ $<$ 0.35 M$_{\odot}$  and reach up to more than 40\% for 
 the few stars with M$_{\star}$ $<$ 0.25 M$_{\odot}$. In a similar
 way, relative errors in radius can be larger than 20\%
 for stars with R$_{\star}$ $<$ 0.35 R$_{\odot}$.
 Relative errors in luminosities are also larger for low-luminosity
 stars, being significantly high (larger than 70\%) for those 
 stars with  $\log (L_{\star}/L_{\odot})$  $<$ -2.
 We point as a possible explanation the fact that relative errors
 in masses obtained from \cite{1993AJ....106..773H} relationship
 tend to be larger at lower masses. 
 
\begin{figure}[!htb]
\centering
\includegraphics[angle=270,scale=0.50]{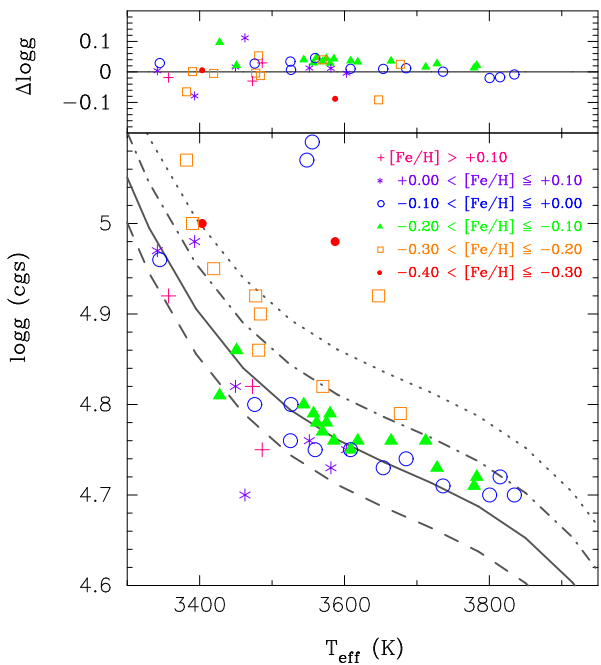}
\caption{
 Logarithmic surface
 gravity 
 as a function of the effective temperature. Stars are plotted
 using different colours and symbols according to their metallicity. Several fits for fixed
 metallicity values are plotted: +0.15 (dashed line), +0.00 (solid line),
 -0.15 (dash-dotted line), and -0.30 (dotted line).
 The upper panel shows the differences between the $\log g$ obtained with our calibrations
 (Equation~\ref{eqn:g}) and those derived from masses and radius.
 Median errors in $\log g$ estimates
 are 0.18 dex (mass-radius derived values)
 and 0.05 dex (values from Equation~\ref{eqn:g}).}
\label{logg_teff_metal_plot}
\end{figure}

  The M$_{\star}$ and R$_{\star}$ 
  versus temperature planes are shown in 
  Figure~\ref{mrg_teff_metal_plot} where the stars are plotted with different colours
  according to their metallicities.
  It can be seen that for a given effective temperature, larger
  stellar metallicities predict larger masses and radii.
  Regarding surface gravity, see Figure~\ref{logg_teff_metal_plot},
  the effect of metallicity tends to be the opposite
  with lower gravities in stars with higher metallicity content. 

\section{Summary}
\label{summary}

 The determination of accurate stellar parameters of low-mass stars is certainly
 a major topic in nowadays astrophysics. 
 This is in part because of their advantages 
 with respect to solar-type stars in the search for small, rocky,
 potentially habitable planets. 
 This fact motivated us to develop a methodology to 
 characterise early M dwarfs using the high-resolution spectra that are being 
 obtained in the current radial velocity exoplanet programmes. 
 We made use of ratios of features as a method to determine effective temperatures,
 and combinations of features and temperature-sensitive ratios to determine metallicities.
 This technique largely applied to solar-type, subgiant, and giant stars had not been extended before
 to the low-mass stars regime, probably because of the difficulty in identifying spectral lines in their
 optical spectra. We also provide empirical calibrations for masses, radii, and gravities
 as a function of effective temperature and metallicity.
 Our main results are as follows:

 \begin{itemize}


 \item  The behaviour of the EW of features
 was studied as a function of the effective temperature and the metallicity.
 The results show that for a significant fraction of the features, 
 $\sim$ 50\%, the EW shows a high anticorrelation with T$_{\rm eff}$, whilst
 the correlations between EW and metallicity are in general weaker.

 \item  Empirical calibrations for the effective temperature were obtained
 using stars with interferometric measurement of their radii from BO12
 as calibrators.  112 ratios of features sensitive to the temperature were calibrated
 providing effective temperatures with  
 typical uncertainties of the order of 70 K.  
 
 \item  In the same way 82 ratios of features were calibrated to derive spectral types.

 \item Stellar metallicities were obtained from  696 combinations of EW of
 individual features
 and temperature-sensitive ratios, with estimated uncertainties in the range of
 0.07-0.10 dex. 
  
 \item We made use of our technique to characterise 53 early M dwarfs which are currently
 being monitored in the HARPS exoplanet search programme. 
 Photometric estimates
 of stellar mass, radius, and surface gravity were used
 to search for possible correlations of these parameters on  T$_{\rm eff}$ and
 [Fe/H].  

 \item We found stellar masses, radii, and surface gravities
 to have a moderate but statistically significant
 correlation with the stellar metallicity, in the sense that at a given effective temperature
 larger metallicities predict slightly larger masses and radii
 whereas, larger gravities are found in stars with 
 lower metallicity content. 

\end{itemize}

 Although high-resolution HARPS and HARPS-N optical spectra were used for this work, a similar
 methodology can be used to derive T$_{\rm eff}$ and [Fe/H] for other instruments/spectral
 ranges. 

\begin{acknowledgements}

  This work was supported by the Italian Ministry of Education,
  University, and Research  through the
  \emph{Premiale HARPS-N} research project under grant
  \emph{Ricerca di pianeti intorno a stelle di piccola massa}.
  M. P. and I. R. acknowledge financial support from the Spanish Ministry
  of Economy and Competitiveness (MINECO) and the \emph{Fondo Europeo de
  Desarrollo Regional} (FEDER) through grants AYA2012-39612-C03-01 and
  ESP2013-48391-C4-1-R. We sincerely appreciate the careful
  reading of the manuscript and the constructive comments of an anonymous
  referee. 

\end{acknowledgements}


\bibliographystyle{aa}
\bibliography{mdwarfs.bib}

\begin{thebibliography}{62}
\expandafter\ifx\csname natexlab\endcsname\relax\def\natexlab#1{#1}\fi

\bibitem[{{Allard} {et~al.}(2012{\natexlab{a}}){Allard}, {Homeier}, \&
  {Freytag}}]{2012RSPTA.370.2765A}
{Allard}, F., {Homeier}, D., \& {Freytag}, B. 2012{\natexlab{a}}, Royal Society
  of London Philosophical Transactions Series A, 370, 2765

\bibitem[{{Allard} {et~al.}(2012{\natexlab{b}}){Allard}, {Homeier}, {Freytag},
  \& {Sharp}}]{2012EAS....57....3A}
{Allard}, F., {Homeier}, D., {Freytag}, B., \& {Sharp}, C.~M.
  2012{\natexlab{b}}, in EAS Publications Series, Vol.~57, EAS Publications
  Series, ed. C.~{Reyl{\'e}}, C.~{Charbonnel}, \& M.~{Schultheis}, 3--43

\bibitem[{{Baraffe} {et~al.}(1998){Baraffe}, {Chabrier}, {Allard}, \&
  {Hauschildt}}]{1998A&A...337..403B}
{Baraffe}, I., {Chabrier}, G., {Allard}, F., \& {Hauschildt}, P.~H. 1998, \aap,
  337, 403

\bibitem[{{Bean} {et~al.}(2006){Bean}, {Benedict}, \&
  {Endl}}]{2006ApJ...653L..65B}
{Bean}, J.~L., {Benedict}, G.~F., \& {Endl}, M. 2006, \apjl, 653, L65

\bibitem[{{Biazzo} {et~al.}(2007){Biazzo}, {Frasca}, {Catalano}, \&
  {Marilli}}]{2007AN....328..938B}
{Biazzo}, K., {Frasca}, A., {Catalano}, S., \& {Marilli}, E. 2007,
  Astronomische Nachrichten, 328, 938

\bibitem[{{Bonfils} {et~al.}(2013){Bonfils}, {Delfosse}, {Udry}, {Forveille},
  {Mayor}, {Perrier}, {Bouchy}, {Gillon}, {Lovis}, {Pepe}, {Queloz}, {Santos},
  {S{\'e}gransan}, \& {Bertaux}}]{2013A&A...549A.109B}
{Bonfils}, X., {Delfosse}, X., {Udry}, S., {et~al.} 2013, \aap, 549, A109

\bibitem[{{Bonfils} {et~al.}(2005){Bonfils}, {Delfosse}, {Udry}, {Santos},
  {Forveille}, \& {S{\'e}gransan}}]{2005A&A...442..635B}
{Bonfils}, X., {Delfosse}, X., {Udry}, S., {et~al.} 2005, \aap, 442, 635

\bibitem[{{Boyajian} {et~al.}(2012){Boyajian}, {von Braun}, {van Belle},
  {McAlister}, {ten Brummelaar}, {Kane}, {Muirhead}, {Jones}, {White},
  {Schaefer}, {Ciardi}, {Henry}, {L{\'o}pez-Morales}, {Ridgway}, {Gies}, {Jao},
  {Rojas-Ayala}, {Parks}, {Sturmann}, {Sturmann}, {Turner}, {Farrington},
  {Goldfinger}, \& {Berger}}]{2012ApJ...757..112B}
{Boyajian}, T.~S., {von Braun}, K., {van Belle}, G., {et~al.} 2012, \apj, 757,
  112

\bibitem[{{Caccin} {et~al.}(2002){Caccin}, {Penza}, \&
  {Gomez}}]{2002A&A...386..286C}
{Caccin}, B., {Penza}, V., \& {Gomez}, M.~T. 2002, \aap, 386, 286

\bibitem[{{Carpenter}(2001)}]{2001AJ....121.2851C}
{Carpenter}, J.~M. 2001, \aj, 121, 2851

\bibitem[{{Casagrande} {et~al.}(2008){Casagrande}, {Flynn}, \&
  {Bessell}}]{2008MNRAS.389..585C}
{Casagrande}, L., {Flynn}, C., \& {Bessell}, M. 2008, \mnras, 389, 585

\bibitem[{{Casagrande} {et~al.}(2006){Casagrande}, {Portinari}, \&
  {Flynn}}]{2006MNRAS.373...13C}
{Casagrande}, L., {Portinari}, L., \& {Flynn}, C. 2006, \mnras, 373, 13

\bibitem[{{Cosentino} {et~al.}(2012){Cosentino}, {Lovis}, {Pepe}, {Collier
  Cameron}, {Latham}, {Molinari}, {Udry}, {Bezawada}, {Black}, {Born},
  {Buchschacher}, {Charbonneau}, {Figueira}, {Fleury}, {Galli}, {Gallie},
  {Gao}, {Ghedina}, {Gonzalez}, {Gonzalez}, {Guerra}, {Henry}, {Horne},
  {Hughes}, {Kelly}, {Lodi}, {Lunney}, {Maire}, {Mayor}, {Micela}, {Ordway},
  {Peacock}, {Phillips}, {Piotto}, {Pollacco}, {Queloz}, {Rice}, {Riverol},
  {Riverol}, {San Juan}, {Sasselov}, {Segransan}, {Sozzetti}, {Sosnowska},
  {Stobie}, {Szentgyorgyi}, {Vick}, \& {Weber}}]{2012SPIE.8446E..1VC}
{Cosentino}, R., {Lovis}, C., {Pepe}, F., {et~al.} 2012, in Society of
  Photo-Optical Instrumentation Engineers (SPIE) Conference Series, Vol. 8446,
  Society of Photo-Optical Instrumentation Engineers (SPIE) Conference Series,
  1

\bibitem[{{Covey} {et~al.}(2007){Covey}, {Ivezi{\'c}}, {Schlegel},
  {Finkbeiner}, {Padmanabhan}, {Lupton}, {Ag{\"u}eros}, {Bochanski}, {Hawley},
  {West}, {Seth}, {Kimball}, {Gogarten}, {Claire}, {Haggard}, {Kaib},
  {Schneider}, \& {Sesar}}]{2007AJ....134.2398C}
{Covey}, K.~R., {Ivezi{\'c}}, {\v Z}., {Schlegel}, D., {et~al.} 2007, \aj, 134,
  2398

\bibitem[{{Covino} {et~al.}(2013){Covino}, {Esposito}, {Barbieri}, {Mancini},
  {Nascimbeni}, {Claudi}, {Desidera}, {Gratton}, {Lanza}, {Sozzetti}, {Biazzo},
  {Affer}, {Gandolfi}, {Munari}, {Pagano}, {Bonomo}, {Collier Cameron},
  {H{\'e}brard}, {Maggio}, {Messina}, {Micela}, {Molinari}, {Pepe}, {Piotto},
  {Ribas}, {Santos}, {Southworth}, {Shkolnik}, {Triaud}, {Bedin}, {Benatti},
  {Boccato}, {Bonavita}, {Borsa}, {Borsato}, {Brown}, {Carolo}, {Ciceri},
  {Cosentino}, {Damasso}, {Faedi}, {Mart{\'{\i}}nez Fiorenzano}, {Latham},
  {Lovis}, {Mordasini}, {Nikolov}, {Poretti}, {Rainer}, {Rebolo L{\'o}pez},
  {Scandariato}, {Silvotti}, {Smareglia}, {Alcal{\'a}}, {Cunial}, {Di
  Fabrizio}, {Di Mauro}, {Giacobbe}, {Granata}, {Harutyunyan}, {Knapic},
  {Lattanzi}, {Leto}, {Lodato}, {Malavolta}, {Marzari}, {Molinaro},
  {Nardiello}, {Pedani}, {Prisinzano}, \& {Turrini}}]{2013A&A...554A..28C}
{Covino}, E., {Esposito}, M., {Barbieri}, M., {et~al.} 2013, \aap, 554, A28

\bibitem[{{Curran}(2014)}]{2014arXiv1411.3816C}
{Curran}, P.~A. 2014, ArXiv e-prints

\bibitem[{{Cutri} {et~al.}(2003){Cutri}, {Skrutskie}, {van Dyk}, {Beichman},
  {Carpenter}, {Chester}, {Cambresy}, {Evans}, {Fowler}, {Gizis}, {Howard},
  {Huchra}, {Jarrett}, {Kopan}, {Kirkpatrick}, {Light}, {Marsh}, {McCallon},
  {Schneider}, {Stiening}, {Sykes}, {Weinberg}, {Wheaton}, {Wheelock}, \&
  {Zacarias}}]{2003yCat.2246....0C}
{Cutri}, R.~M., {Skrutskie}, M.~F., {van Dyk}, S., {et~al.} 2003, VizieR Online
  Data Catalog, 2246, 0

\bibitem[{{Datson} {et~al.}(2012){Datson}, {Flynn}, \&
  {Portinari}}]{2012MNRAS.426..484D}
{Datson}, J., {Flynn}, C., \& {Portinari}, L. 2012, \mnras, 426, 484

\bibitem[{{Datson} {et~al.}(2014{\natexlab{a}}){Datson}, {Flynn}, \&
  {Portinari}}]{2014MNRAS.439.1028D}
{Datson}, J., {Flynn}, C., \& {Portinari}, L. 2014{\natexlab{a}}, \mnras, 439,
  1028

\bibitem[{{Datson} {et~al.}(2014{\natexlab{b}}){Datson}, {Flynn}, \&
  {Portinari}}]{2014arXiv1412.8168D}
{Datson}, J., {Flynn}, C., \& {Portinari}, L. 2014{\natexlab{b}}, ArXiv
  e-prints

\bibitem[{{Dhital} {et~al.}(2012){Dhital}, {West}, {Stassun}, {Bochanski},
  {Massey}, \& {Bastien}}]{2012AJ....143...67D}
{Dhital}, S., {West}, A.~A., {Stassun}, K.~G., {et~al.} 2012, \aj, 143, 67

\bibitem[{{Dotter} {et~al.}(2008){Dotter}, {Chaboyer}, {Jevremovi{\'c}},
  {Kostov}, {Baron}, \& {Ferguson}}]{2008ApJS..178...89D}
{Dotter}, A., {Chaboyer}, B., {Jevremovi{\'c}}, D., {et~al.} 2008, \apjs, 178,
  89

\bibitem[{{Dressing} \& {Charbonneau}(2013)}]{2013ApJ...767...95D}
{Dressing}, C.~D. \& {Charbonneau}, D. 2013, \apj, 767, 95

\bibitem[{{Gaidos} {et~al.}(2014){Gaidos}, {Mann}, {L{\'e}pine}, {Buccino},
  {James}, {Ansdell}, {Petrucci}, {Mauas}, \& {Hilton}}]{2014MNRAS.443.2561G}
{Gaidos}, E., {Mann}, A.~W., {L{\'e}pine}, S., {et~al.} 2014, \mnras, 443, 2561

\bibitem[{{Gliese} \& {Jahrei{\ss}}(1991)}]{1991adc..rept.....G}
{Gliese}, W. \& {Jahrei{\ss}}, H. 1991, {Preliminary Version of the Third
  Catalogue of Nearby Stars}, Tech. rep.

\bibitem[{{Gray}(1989)}]{1989ApJ...347.1021G}
{Gray}, D.~F. 1989, \apj, 347, 1021

\bibitem[{{Gray}(1994)}]{1994PASP..106.1248G}
{Gray}, D.~F. 1994, \pasp, 106, 1248

\bibitem[{{Gray} \& {Johanson}(1991)}]{1991PASP..103..439G}
{Gray}, D.~F. \& {Johanson}, H.~L. 1991, \pasp, 103, 439

\bibitem[{{Hartman} {et~al.}(2014){Hartman}, {Bayliss}, {Brahm}, {Bakos},
  {Mancini}, {Jord{\'a}n}, {Penev}, {Rabus}, {Zhou}, {Butler}, {Espinoza}, {de
  Val-Borro}, {Bhatti}, {Csubry}, {Ciceri}, {Henning}, {Schmidt}, {Arriagada},
  {Shectman}, {Crane}, {Thompson}, {Suc}, {Cs{\'a}k}, {Tan}, {Noyes},
  {L{\'a}z{\'a}r}, {Papp}, \& {S{\'a}ri}}]{2014arXiv1408.1758H}
{Hartman}, J.~D., {Bayliss}, D., {Brahm}, R., {et~al.} 2014, ArXiv e-prints

\bibitem[{{Henry} {et~al.}(1994){Henry}, {Kirkpatrick}, \&
  {Simons}}]{1994AJ....108.1437H}
{Henry}, T.~J., {Kirkpatrick}, J.~D., \& {Simons}, D.~A. 1994, \aj, 108, 1437

\bibitem[{{Henry} \& {McCarthy}(1993)}]{1993AJ....106..773H}
{Henry}, T.~J. \& {McCarthy}, Jr., D.~W. 1993, \aj, 106, 773

\bibitem[{{Johnson} \& {Apps}(2009)}]{2009ApJ...699..933J}
{Johnson}, J.~A. \& {Apps}, K. 2009, \apj, 699, 933

\bibitem[{{Kenyon} \& {Hartmann}(1995)}]{1995ApJS..101..117K}
{Kenyon}, S.~J. \& {Hartmann}, L. 1995, \apjs, 101, 117

\bibitem[{{Kirkpatrick} {et~al.}(1991){Kirkpatrick}, {Henry}, \&
  {McCarthy}}]{1991ApJS...77..417K}
{Kirkpatrick}, J.~D., {Henry}, T.~J., \& {McCarthy}, Jr., D.~W. 1991, \apjs,
  77, 417

\bibitem[{{Kovtyukh} \& {Gorlova}(2000)}]{2000A&A...358..587K}
{Kovtyukh}, V.~V. \& {Gorlova}, N.~I. 2000, \aap, 358, 587

\bibitem[{{Kovtyukh} {et~al.}(2003){Kovtyukh}, {Soubiran}, {Belik}, \&
  {Gorlova}}]{2003A&A...411..559K}
{Kovtyukh}, V.~V., {Soubiran}, C., {Belik}, S.~I., \& {Gorlova}, N.~I. 2003,
  \aap, 411, 559

\bibitem[{{L{\'e}pine} {et~al.}(2013){L{\'e}pine}, {Hilton}, {Mann}, {Wilde},
  {Rojas-Ayala}, {Cruz}, \& {Gaidos}}]{2013AJ....145..102L}
{L{\'e}pine}, S., {Hilton}, E.~J., {Mann}, A.~W., {et~al.} 2013, \aj, 145, 102

\bibitem[{{L{\'e}pine} {et~al.}(2007){L{\'e}pine}, {Rich}, \&
  {Shara}}]{2007ApJ...669.1235L}
{L{\'e}pine}, S., {Rich}, R.~M., \& {Shara}, M.~M. 2007, \apj, 669, 1235

\bibitem[{{Lupton}(1993)}]{1993stp..book.....L}
{Lupton}, R. 1993, {Statistics in theory and practice}

\bibitem[{{Maness} {et~al.}(2007){Maness}, {Marcy}, {Ford}, {Hauschildt},
  {Shreve}, {Basri}, {Butler}, \& {Vogt}}]{2007PASP..119...90M}
{Maness}, H.~L., {Marcy}, G.~W., {Ford}, E.~B., {et~al.} 2007, \pasp, 119, 90

\bibitem[{{Mann} {et~al.}(2013{\natexlab{a}}){Mann}, {Brewer}, {Gaidos},
  {L{\'e}pine}, \& {Hilton}}]{2013AJ....145...52M}
{Mann}, A.~W., {Brewer}, J.~M., {Gaidos}, E., {L{\'e}pine}, S., \& {Hilton},
  E.~J. 2013{\natexlab{a}}, \aj, 145, 52

\bibitem[{{Mann} {et~al.}(2015){Mann}, {Feiden}, {Gaidos}, \&
  {Boyajian}}]{2015arXiv150101635M}
{Mann}, A.~W., {Feiden}, G.~A., {Gaidos}, E., \& {Boyajian}, T. 2015, ArXiv
  e-prints

\bibitem[{{Mann} {et~al.}(2013{\natexlab{b}}){Mann}, {Gaidos}, \&
  {Ansdell}}]{2013ApJ...779..188M}
{Mann}, A.~W., {Gaidos}, E., \& {Ansdell}, M. 2013{\natexlab{b}}, \apj, 779,
  188

\bibitem[{{Markwardt}(2009)}]{2009ASPC..411..251M}
{Markwardt}, C.~B. 2009, in Astronomical Society of the Pacific Conference
  Series, Vol. 411, Astronomical Data Analysis Software and Systems XVIII, ed.
  D.~A. {Bohlender}, D.~{Durand}, \& P.~{Dowler}, 251

\bibitem[{{Mayor} {et~al.}(2003){Mayor}, {Pepe}, {Queloz}, {Bouchy},
  {Rupprecht}, {Lo Curto}, {Avila}, {Benz}, {Bertaux}, {Bonfils}, {Dall},
  {Dekker}, {Delabre}, {Eckert}, {Fleury}, {Gilliotte}, {Gojak}, {Guzman},
  {Kohler}, {Lizon}, {Longinotti}, {Lovis}, {Megevand}, {Pasquini}, {Reyes},
  {Sivan}, {Sosnowska}, {Soto}, {Udry}, {van Kesteren}, {Weber}, \&
  {Weilenmann}}]{2003Msngr.114...20M}
{Mayor}, M., {Pepe}, F., {Queloz}, D., {et~al.} 2003, The Messenger, 114, 20

\bibitem[{{Montes} {et~al.}(2007){Montes}, {Mart{\'{\i}}nez-Arn{\'a}iz},
  {Maldonado}, {Roa-Llamazares}, {L{\'o}pez-Santiago}, {Crespo-Chac{\'o}n}, \&
  {Solano}}]{2007HiA....14..598M}
{Montes}, D., {Mart{\'{\i}}nez-Arn{\'a}iz}, R.~M., {Maldonado}, J., {et~al.}
  2007, Highlights of Astronomy, 14, 598

\bibitem[{{Neves} {et~al.}(2012){Neves}, {Bonfils}, {Santos}, {Delfosse},
  {Forveille}, {Allard}, {Nat{\'a}rio}, {Fernandes}, \&
  {Udry}}]{2012A&A...538A..25N}
{Neves}, V., {Bonfils}, X., {Santos}, N.~C., {et~al.} 2012, \aap, 538, A25

\bibitem[{{Neves} {et~al.}(2014){Neves}, {Bonfils}, {Santos}, {Delfosse},
  {Forveille}, {Allard}, \& {Udry}}]{2014A&A...568A.121N}
{Neves}, V., {Bonfils}, X., {Santos}, N.~C., {et~al.} 2014, \aap, 568, A121

\bibitem[{{Newton} {et~al.}(2014{\natexlab{a}}){Newton}, {Charbonneau},
  {Irwin}, {Berta-Thompson}, {Rojas-Ayala}, {Covey}, \&
  {Lloyd}}]{2014AJ....147...20N}
{Newton}, E.~R., {Charbonneau}, D., {Irwin}, J., {et~al.} 2014{\natexlab{a}},
  \aj, 147, 20

\bibitem[{{Newton} {et~al.}(2014{\natexlab{b}}){Newton}, {Charbonneau},
  {Irwin}, \& {Mann}}]{2014arXiv1412.2758N}
{Newton}, E.~R., {Charbonneau}, D., {Irwin}, J., \& {Mann}, A.~W.
  2014{\natexlab{b}}, ArXiv e-prints

\bibitem[{{{\"O}nehag} {et~al.}(2012){{\"O}nehag}, {Heiter}, {Gustafsson},
  {Piskunov}, {Plez}, \& {Reiners}}]{2012A&A...542A..33O}
{{\"O}nehag}, A., {Heiter}, U., {Gustafsson}, B., {et~al.} 2012, \aap, 542, A33

\bibitem[{{Rajpurohit} {et~al.}(2014){Rajpurohit}, {Reyl{\'e}}, {Allard},
  {Scholz}, {Homeier}, {Schultheis}, \& {Bayo}}]{2014A&A...564A..90R}
{Rajpurohit}, A.~S., {Reyl{\'e}}, C., {Allard}, F., {et~al.} 2014, \aap, 564,
  A90

\bibitem[{{Rojas-Ayala} {et~al.}(2012){Rojas-Ayala}, {Covey}, {Muirhead}, \&
  {Lloyd}}]{2012ApJ...748...93R}
{Rojas-Ayala}, B., {Covey}, K.~R., {Muirhead}, P.~S., \& {Lloyd}, J.~P. 2012,
  \apj, 748, 93

\bibitem[{{Schlaufman} \& {Laughlin}(2010)}]{2010A&A...519A.105S}
{Schlaufman}, K.~C. \& {Laughlin}, G. 2010, \aap, 519, A105

\bibitem[{{S{\'e}gransan} {et~al.}(2003){S{\'e}gransan}, {Kervella},
  {Forveille}, \& {Queloz}}]{2003A&A...397L...5S}
{S{\'e}gransan}, D., {Kervella}, P., {Forveille}, T., \& {Queloz}, D. 2003,
  \aap, 397, L5

\bibitem[{{Sousa} {et~al.}(2010){Sousa}, {Alapini}, {Israelian}, \&
  {Santos}}]{2010A&A...512A..13S}
{Sousa}, S.~G., {Alapini}, A., {Israelian}, G., \& {Santos}, N.~C. 2010, \aap,
  512, A13

\bibitem[{{Sozzetti} {et~al.}(2013){Sozzetti}, {Bernagozzi}, {Bertolini},
  {Calcidese}, {Carbognani}, {Cenadelli}, {Christille}, {Damasso}, {Giacobbe},
  {Lanteri}, {Lattanzi}, \& {Smart}}]{2013EPJWC..4703006S}
{Sozzetti}, A., {Bernagozzi}, A., {Bertolini}, E., {et~al.} 2013, in European
  Physical Journal Web of Conferences, Vol.~47, European Physical Journal Web
  of Conferences, 3006

\bibitem[{{Strassmeier} \& {Schordan}(2000)}]{2000AN....321..277S}
{Strassmeier}, K.~G. \& {Schordan}, P. 2000, Astronomische Nachrichten, 321,
  277

\bibitem[{{Terrien} {et~al.}(2012){Terrien}, {Mahadevan}, {Bender},
  {Deshpande}, {Ramsey}, \& {Bochanski}}]{2012ApJ...747L..38T}
{Terrien}, R.~C., {Mahadevan}, S., {Bender}, C.~F., {et~al.} 2012, \apjl, 747,
  L38

\bibitem[{{von Braun} {et~al.}(2014){von Braun}, {Boyajian}, {van Belle},
  {Kane}, {Jones}, {Farrington}, {Schaefer}, {Vargas}, {Scott}, {ten
  Brummelaar}, {Kephart}, {Gies}, {Ciardi}, {L{\'o}pez-Morales}, {Mazingue},
  {McAlister}, {Ridgway}, {Goldfinger}, {Turner}, \&
  {Sturmann}}]{2014MNRAS.438.2413V}
{von Braun}, K., {Boyajian}, T.~S., {van Belle}, G.~T., {et~al.} 2014, \mnras,
  438, 2413

\bibitem[{{West} {et~al.}(2011){West}, {Morgan}, {Bochanski}, {Andersen},
  {Bell}, {Kowalski}, {Davenport}, {Hawley}, {Schmidt}, {Bernat}, {Hilton},
  {Muirhead}, {Covey}, {Rojas-Ayala}, {Schlawin}, {Gooding}, {Schluns},
  {Dhital}, {Pineda}, \& {Jones}}]{2011AJ....141...97W}
{West}, A.~A., {Morgan}, D.~P., {Bochanski}, J.~J., {et~al.} 2011, \aj, 141, 97

\bibitem[{{Woolf} \& {Wallerstein}(2005)}]{2005MNRAS.356..963W}
{Woolf}, V.~M. \& {Wallerstein}, G. 2005, \mnras, 356, 963

\end{thebibliography}

\Online
\section*{Online material}
\label{tables}

  Table~\ref{parameters_table_full} lists all the stars analyzed in this work.
  The table provides:
  Star identifier (column 1);
  effective temperature in kelvin (column 2);
  spectral type (column 3);
  stellar metallicity in dex (column 4);
  logarithm of the surface gravity, $\log g$, in cms$^{\rm -2}$ (column 5);
  stellar mass in solar units (column 6);
  stellar radius in solar units (column 7);
  and  stellar luminosity, $\log (L_{\star}/L_{\odot})$ (column 8).
  Each measured quantity is accompanied by its corresponding uncertainty.

\onllongtab{
\begin{longtable}{lccccccc}
\caption{
 Stellar parameters 
 for the stars studied in this work.
  }\label{parameters_table_full}\\
\hline
\hline
 Star   & T$_{\rm eff}$ &  Sp-Type   & [Fe/H]   & $\log g$   & M$_{\star}$     & R$_{\star}$    &  $\log (L_{\star}/L_{\odot})$   \\
        &   (K)         &            & (dex)    &   (cgs)    & (M$_{\odot}$)   & (R$_{\odot}$)  &      \\
 (1)    &   (2)         &  (3)       & (4)      &   (5)      & (6)             &  (7)           & (8)  \\
\hline
\endfirsthead
\caption{Continued.} \\
\hline
 Star   & T$_{\rm eff}$ &  Sp-Type   & [Fe/H]   & $\log g$    & M$_{\star}$     & R$_{\star}$    &  $\log (L_{\star}/L_{\odot})$   \\
        &   (K)         &            & (dex)    &   (cgs)     & (M$_{\odot}$)   & (R$_{\odot}$)  &      \\
 (1)    &   (2)         &  (3)       & (4)      &   (5)       & (6)             &  (7)           & (8)   \\
\hline
\endhead
\hline
\endfoot
\hline
\endlastfoot
Gl1	&	3482	$\pm$	68	&	M2.5	&	-0.27	$\pm$	0.09	&	4.91	$\pm$	0.06	&	0.33	$\pm$	21.39	\%	&	0.33	$\pm$	18.84	\%	&	-1.834	$\pm$	38.49	\%	\\
Gl87	&	3562	$\pm$	68	&	M2	&	-0.16	$\pm$	0.09	&	4.83	$\pm$	0.05	&	0.43	$\pm$	13.22	\%	&	0.42	$\pm$	11.99	\%	&	-1.590	$\pm$	25.18	\%	\\
Gl176	&	3603	$\pm$	68	&	M2	&	0.03	$\pm$	0.09	&	4.75	$\pm$	0.04	&	0.52	$\pm$	10.07	\%	&	0.51	$\pm$	9.31	\%	&	-1.409	$\pm$	20.08	\%	\\
Gl191	&	3587	$\pm$	68	&	M0	&	-0.39	$\pm$	0.09	&	4.89	$\pm$	0.05	&	0.35	$\pm$	15.26	\%	&	0.35	$\pm$	13.64	\%	&	-1.728	$\pm$	28.31	\%	\\
Gl205	&	3800	$\pm$	68	&	M1.5	&	0.00	$\pm$	0.09	&	4.68	$\pm$	0.05	&	0.60	$\pm$	9.35	\%	&	0.58	$\pm$	8.99	\%	&	-1.194	$\pm$	19.36	\%	\\
Gl229	&	3779	$\pm$	69	&	M1	&	-0.10	$\pm$	0.09	&	4.72	$\pm$	0.05	&	0.55	$\pm$	9.81	\%	&	0.54	$\pm$	9.34	\%	&	-1.276	$\pm$	20.04	\%	\\
HIP31293	&	3526	$\pm$	68	&	M3	&	-0.04	$\pm$	0.09	&	4.81	$\pm$	0.05	&	0.45	$\pm$	13.63	\%	&	0.44	$\pm$	12.39	\%	&	-1.564	$\pm$	25.95	\%	\\
Gl250B	&	3557	$\pm$	68	&	M2.5	&	-0.13	$\pm$	0.09	&	4.82	$\pm$	0.05	&	0.44	$\pm$	13.08	\%	&	0.43	$\pm$	11.88	\%	&	-1.576	$\pm$	24.97	\%	\\
Gl273	&	3342	$\pm$	69	&	M4	&	0.01	$\pm$	0.09	&	4.97	$\pm$	0.11	&	0.27	$\pm$	44.09	\%	&	0.28	$\pm$	37.91	\%	&	-2.055	$\pm$	76.27	\%	\\
GJ2066	&	3575	$\pm$	68	&	M2	&	-0.17	$\pm$	0.09	&	4.82	$\pm$	0.05	&	0.43	$\pm$	12.75	\%	&	0.43	$\pm$	11.59	\%	&	-1.577	$\pm$	24.40	\%	\\
Gl341	&	3783	$\pm$	69	&	M0.5	&	-0.16	$\pm$	0.09	&	4.74	$\pm$	0.05	&	0.53	$\pm$	10.27	\%	&	0.52	$\pm$	9.75	\%	&	-1.305	$\pm$	20.81	\%	\\
Gl357	&	3477	$\pm$	68	&	M2.5	&	-0.27	$\pm$	0.09	&	4.92	$\pm$	0.06	&	0.32	$\pm$	21.83	\%	&	0.33	$\pm$	19.21	\%	&	-1.844	$\pm$	39.21	\%	\\
Gl358	&	3450	$\pm$	68	&	M3	&	0.04	$\pm$	0.09	&	4.84	$\pm$	0.07	&	0.42	$\pm$	18.71	\%	&	0.42	$\pm$	16.86	\%	&	-1.660	$\pm$	34.63	\%	\\
Gl367	&	3559	$\pm$	68	&	M2.5	&	-0.06	$\pm$	0.09	&	4.80	$\pm$	0.05	&	0.46	$\pm$	12.22	\%	&	0.46	$\pm$	11.15	\%	&	-1.525	$\pm$	23.59	\%	\\
Gl382	&	3653	$\pm$	68	&	M2	&	-0.01	$\pm$	0.09	&	4.74	$\pm$	0.04	&	0.53	$\pm$	9.48	\%	&	0.52	$\pm$	8.82	\%	&	-1.372	$\pm$	19.15	\%	\\
Gl388	&	3473	$\pm$	68	&	M3.5	&	0.12	$\pm$	0.10	&	4.79	$\pm$	0.07	&	0.47	$\pm$	15.76	\%	&	0.46	$\pm$	14.36	\%	&	-1.555	$\pm$	29.77	\%	\\
Gl393	&	3544	$\pm$	68	&	M2	&	-0.17	$\pm$	0.09	&	4.84	$\pm$	0.05	&	0.41	$\pm$	14.23	\%	&	0.41	$\pm$	12.86	\%	&	-1.626	$\pm$	26.84	\%	\\
Gl413.1	&	3570	$\pm$	68	&	M2	&	-0.10	$\pm$	0.09	&	4.80	$\pm$	0.05	&	0.45	$\pm$	12.21	\%	&	0.45	$\pm$	11.14	\%	&	-1.536	$\pm$	23.56	\%	\\
Gl433	&	3618	$\pm$	68	&	M1.5	&	-0.13	$\pm$	0.09	&	4.79	$\pm$	0.04	&	0.47	$\pm$	11.02	\%	&	0.46	$\pm$	10.13	\%	&	-1.490	$\pm$	21.61	\%	\\
Gl438	&	3647	$\pm$	68	&	M1	&	-0.27	$\pm$	0.09	&	4.83	$\pm$	0.04	&	0.43	$\pm$	11.78	\%	&	0.42	$\pm$	10.78	\%	&	-1.548	$\pm$	22.82	\%	\\
Gl447	&	3382	$\pm$	68	&	M4	&	-0.25	$\pm$	0.09	&	5.00	$\pm$	0.09	&	0.23	$\pm$	44.05	\%	&	0.24	$\pm$	36.99	\%	&	-2.156	$\pm$	74.41	\%	\\
Gl465	&	3403	$\pm$	69	&	M2.5	&	-0.33	$\pm$	0.09	&	5.01	$\pm$	0.09	&	0.23	$\pm$	41.42	\%	&	0.24	$\pm$	34.72	\%	&	-2.154	$\pm$	69.91	\%	\\
Gl479	&	3476	$\pm$	68	&	M3	&	0.00	$\pm$	0.09	&	4.83	$\pm$	0.06	&	0.43	$\pm$	16.71	\%	&	0.42	$\pm$	15.10	\%	&	-1.630	$\pm$	31.20	\%	\\
Gl514	&	3728	$\pm$	68	&	M1	&	-0.14	$\pm$	0.09	&	4.76	$\pm$	0.04	&	0.51	$\pm$	9.92	\%	&	0.50	$\pm$	9.30	\%	&	-1.362	$\pm$	20.00	\%	\\
Gl526	&	3609	$\pm$	68	&	M2	&	-0.10	$\pm$	0.09	&	4.79	$\pm$	0.04	&	0.47	$\pm$	10.99	\%	&	0.47	$\pm$	10.10	\%	&	-1.483	$\pm$	21.57	\%	\\
Gl536	&	3685	$\pm$	68	&	M1	&	-0.08	$\pm$	0.09	&	4.75	$\pm$	0.04	&	0.52	$\pm$	9.66	\%	&	0.50	$\pm$	9.00	\%	&	-1.377	$\pm$	19.47	\%	\\
Gl551	&	3555	$\pm$	68	&	M4.5	&	-0.03	$\pm$	0.09	&	4.79	$\pm$	0.05	&	0.47	$\pm$	12.04	\%	&	0.46	$\pm$	11.01	\%	&	-1.511	$\pm$	23.29	\%	\\
Gl569A	&	3608	$\pm$	68	&	M2	&	-0.02	$\pm$	0.09	&	4.76	$\pm$	0.04	&	0.50	$\pm$	10.32	\%	&	0.49	$\pm$	9.52	\%	&	-1.433	$\pm$	20.48	\%	\\
Gl581	&	3419	$\pm$	68	&	M3	&	-0.20	$\pm$	0.09	&	4.95	$\pm$	0.08	&	0.29	$\pm$	29.73	\%	&	0.30	$\pm$	25.84	\%	&	-1.948	$\pm$	52.29	\%	\\
Gl588	&	3525	$\pm$	68	&	M3	&	0.00	$\pm$	0.09	&	4.79	$\pm$	0.05	&	0.46	$\pm$	13.21	\%	&	0.46	$\pm$	12.04	\%	&	-1.539	$\pm$	25.28	\%	\\
Gl618A	&	3451	$\pm$	68	&	M3	&	-0.10	$\pm$	0.09	&	4.88	$\pm$	0.07	&	0.37	$\pm$	21.30	\%	&	0.37	$\pm$	18.96	\%	&	-1.764	$\pm$	38.74	\%	\\
Gl628	&	3345	$\pm$	69	&	M3.5	&	-0.05	$\pm$	0.09	&	4.99	$\pm$	0.11	&	0.25	$\pm$	46.49	\%	&	0.26	$\pm$	39.59	\%	&	-2.104	$\pm$	79.61	\%	\\
GJ644A	&	3463	$\pm$	68	&	M2.5	&	0.08	$\pm$	0.10	&	4.81	$\pm$	0.07	&	0.45	$\pm$	17.19	\%	&	0.44	$\pm$	15.59	\%	&	-1.603	$\pm$	32.17	\%	\\
Gl667C$^{\star}$	&	3572	$\pm$	68	&	M1	&				&				&					&					&					\\
Gl674	&	3484	$\pm$	68	&	M2.5	&	-0.20	$\pm$	0.09	&	4.89	$\pm$	0.06	&	0.36	$\pm$	19.49	\%	&	0.36	$\pm$	17.32	\%	&	-1.767	$\pm$	35.52	\%	\\
Gl678.1A	&	3815	$\pm$	69	&	M0	&	-0.09	$\pm$	0.09	&	4.70	$\pm$	0.05	&	0.58	$\pm$	10.13	\%	&	0.56	$\pm$	9.73	\%	&	-1.222	$\pm$	20.77	\%	\\
Gl680	&	3585	$\pm$	68	&	M2	&	-0.12	$\pm$	0.09	&	4.80	$\pm$	0.05	&	0.45	$\pm$	11.84	\%	&	0.45	$\pm$	10.82	\%	&	-1.527	$\pm$	22.93	\%	\\
Gl682	&	3393	$\pm$	69	&	M4	&	0.02	$\pm$	0.09	&	4.90	$\pm$	0.09	&	0.35	$\pm$	28.14	\%	&	0.35	$\pm$	24.92	\%	&	-1.834	$\pm$	50.50	\%	\\
Gl686	&	3677	$\pm$	68	&	M1	&	-0.26	$\pm$	0.09	&	4.81	$\pm$	0.04	&	0.44	$\pm$	11.23	\%	&	0.44	$\pm$	10.34	\%	&	-1.502	$\pm$	21.97	\%	\\
Gl693	&	3390	$\pm$	68	&	M3	&	-0.27	$\pm$	0.09	&	5.00	$\pm$	0.09	&	0.23	$\pm$	42.13	\%	&	0.25	$\pm$	35.45	\%	&	-2.141	$\pm$	71.36	\%	\\
Gl701	&	3664	$\pm$	68	&	M1	&	-0.19	$\pm$	0.09	&	4.80	$\pm$	0.04	&	0.46	$\pm$	10.75	\%	&	0.46	$\pm$	9.92	\%	&	-1.471	$\pm$	21.19	\%	\\
Gl729	&	3548	$\pm$	68	&	M3.5	&	-0.06	$\pm$	0.10	&	4.80	$\pm$	0.05	&	0.46	$\pm$	13.20	\%	&	0.45	$\pm$	12.04	\%	&	-1.542	$\pm$	25.27	\%	\\
Gl752A	&	3551	$\pm$	68	&	M3	&	0.02	$\pm$	0.09	&	4.77	$\pm$	0.05	&	0.49	$\pm$	11.78	\%	&	0.48	$\pm$	10.79	\%	&	-1.485	$\pm$	22.91	\%	\\
Gl803$^{\star}$	&	3628	$\pm$	68	&	M0.5	&				&				&					&					&					\\
Gl832	&	3580	$\pm$	68	&	M2	&	-0.16	$\pm$	0.09	&	4.82	$\pm$	0.05	&	0.44	$\pm$	12.45	\%	&	0.43	$\pm$	11.33	\%	&	-1.562	$\pm$	23.91	\%	\\
Gl846	&	3835	$\pm$	69	&	M0	&	-0.09	$\pm$	0.09	&	4.69	$\pm$	0.05	&	0.59	$\pm$	10.43	\%	&	0.57	$\pm$	10.07	\%	&	-1.194	$\pm$	21.39	\%	\\
Gl849	&	3486	$\pm$	69	&	M3.5	&	0.12	$\pm$	0.09	&	4.78	$\pm$	0.06	&	0.48	$\pm$	14.39	\%	&	0.47	$\pm$	13.13	\%	&	-1.529	$\pm$	27.41	\%	\\
Gl876	&	3357	$\pm$	68	&	M4	&	0.16	$\pm$	0.09	&	4.90	$\pm$	0.10	&	0.35	$\pm$	31.71	\%	&	0.35	$\pm$	28.14	\%	&	-1.845	$\pm$	56.86	\%	\\
Gl877	&	3428	$\pm$	68	&	M3	&	-0.11	$\pm$	0.09	&	4.91	$\pm$	0.08	&	0.34	$\pm$	25.11	\%	&	0.34	$\pm$	22.17	\%	&	-1.838	$\pm$	45.06	\%	\\
Gl880	&	3736	$\pm$	68	&	M1.5	&	-0.01	$\pm$	0.09	&	4.71	$\pm$	0.04	&	0.57	$\pm$	9.04	\%	&	0.55	$\pm$	8.55	\%	&	-1.278	$\pm$	18.59	\%	\\
Gl887	&	3712	$\pm$	68	&	M1	&	-0.18	$\pm$	0.09	&	4.78	$\pm$	0.04	&	0.49	$\pm$	10.26	\%	&	0.48	$\pm$	9.57	\%	&	-1.406	$\pm$	20.50	\%	\\
Gl908	&	3570	$\pm$	68	&	M1.5	&	-0.27	$\pm$	0.09	&	4.86	$\pm$	0.05	&	0.39	$\pm$	14.23	\%	&	0.39	$\pm$	12.82	\%	&	-1.658	$\pm$	26.75	\%	\\
LTT9759	&	3581	$\pm$	68	&	M2.5	&	0.07	$\pm$	0.09	&	4.74	$\pm$	0.05	&	0.52	$\pm$	10.33	\%	&	0.51	$\pm$	9.54	\%	&	-1.413	$\pm$	20.54	\%	\\
\end{longtable}
\tablefoot{$^{\star}$ The star falls out of the range of applicability of our metallicity calibrations.} 
}


\end{document}